\documentclass[aps,prd,superscriptaddress,notitlepage,nofootinbib,noeprint]{revtex4-2}
\usepackage{amsmath,amssymb,amsfonts,amsbsy}
\usepackage{xcolor}
\usepackage{graphicx}
\usepackage{multirow}
\usepackage[normalem]{ulem} 

\begin{document}
\title{The 
light-by-light
contribution to the muon anomalous magnetic moment from the axial-vector mesons exchanges within the nonlocal quark model}

\author{A.E. Radzhabov}
\affiliation{Matrosov Institute for System Dynamics and Control Theory SB RAS, 664033, Irkutsk, Russia}
\author{A.S. Zhevlakov}
\affiliation{Matrosov Institute for System Dynamics and Control Theory SB RAS, 664033, Irkutsk, Russia}
\affiliation{Joint Institute of Nuclear Research, BLTP,141980, Moscow region, Dubna, Russia}
\author{A.P. Martynenko}
\affiliation{Samara University, 443086, Samara, Russia}
\author{F.A. Martynenko}
\affiliation{Samara University, 443086, Samara, Russia}

\begin{abstract}
The contribution of axial-vector mesons to the muon's anomalous magnetic moment through a light-by-light process is considered within a nonlocal quark model. The model is based on a four-quark interaction with scalar--pseudoscalar and vector--axial-vector sectors. While the transverse component of the axial-vector corresponds to a spin-1 particle, the unphysical longitudinal component is mixed with a pseudoscalar meson. 
The model parameters are re-fitted to the pion properties in the presence of  $\pi-a_1$ mixing.
The obtained estimation for the light-by-light contribution of $a_1+f_1$ mesons  is $(3.6\pm1.8)\cdot 10^{-11}$. 

\end{abstract}
\maketitle
\section{Introduction}
\label{intro}
The anomalous magnetic moments  (AMM) $a=(g-2)/2$ of the electron and muon are measured with unprecedented accuracy for elementary particle physics. 
The experimental value for the electron anomaly measured with a one-electron quantum cyclotron is (in units of $10^{-11}$)\footnote{Units $10^{-11}$ will be used through the paper.} 
\cite{Hanneke:2008tm} 
\begin{align}
	a_e^{\mathrm{exp}}&=115 965 218.073\pm0.028.
\end{align}
The value for $a_\mu=(g-2)/2$ has been measured in experiments at  
Brookhaven National Laboratory \cite{Bennett:2006fi} and Fermilab \cite{Muong-2:2021ojo}, and the weighted average is 
\begin{align}
	a_\mu^{\mathrm{BNL}}&= 116 592 089\pm63  , \notag \\
	a_\mu^{\mathrm{FNAL}}&=116 592 040\pm54  ,  \\
	a_\mu^{\mathrm{exp}} &=116 592 061\pm41  . \notag
\end{align}
On the other hand, the magnetic moments of the leptons can be estimated within the framework of the Standard Model, see e.g. \cite{Aoyama:2019ryr} and \cite{Aoyama:2020ynm}. 
While the electron anomaly is mostly due to electromagnetic interactions (the hadronic contribution to electron anomaly is estimated to be only $0.1693 \cdot 10^{-11}$ \cite{Aoyama:2019ryr}), the muon anomaly presents a  challenge for theorists, since it is more sensitive to strong, weak and possible new physics contributions. 
The difference between the experimental measurement and theoretical prediction of the muon magnetic anomaly has long been an excitement to physicists as a possible hint of new physics beyond the Standard Model (SM) which can be observed even at low energies with high precision experiment.  
The difference between the experimental measurement and theoretical prediction of the muon magnetic anomaly is estimated to be \cite{ParticleDataGroup:2022pth}
\begin{align}
a_\mu^{\mathrm{exp}}-a_\mu^{\mathrm{SM}} = 251\pm41 \pm 43. 
\end{align}

In the SM, $a_\mu$ receives contributions from electromagnetic, strong and weak interactions, 
the values quoted for them in \cite{Colangelo:2022jxc} are
\begin{align}
a_\mu^{\mathrm{QED}}&= 116 584 718.931\pm0.104,\notag\\
a_\mu^{\mathrm{HVP}} + a_\mu^{\mathrm{LbL}} &= (6845\pm40) + (92\pm 18), \\
a_\mu^{\mathrm{EW}}&= 153.6\pm1.0,\notag
\end{align}
respectively.
The theoretical understanding of $a_\mu$ is limited mostly due to the strong sector. This is due to the non-perturbative nature of quantum chromodynamics (QCD), i.e. at low energies the strong coupling constant is not a small parameter.
Instead, the data driven approach to the leading order strong contribution, hadronic vacuum polarization (HVP), has been used for a long time, based on the experimental data of total cross section of electron-positron annihilation to hadrons \cite{BouchiatClaude1961}. The recent results can be found in \cite{Davier:2019can,Keshavarzi:2019abf,Colangelo:2018mtw,Hoferichter:2019mqg}.
Due to the necessity of including the isospin breaking effects, the measurement precision of hadronic decays of $\tau$-leptons cannot compete with that of $e^+e^-$.
The ab-initio calculations in the framework of lattice QCD have now reached a level competitive to the data driven approach. The sub-percent uncertainty level is reported by the BMW collaboration \cite{Borsanyi:2020mff} which is however in $\sim 2\sigma$ tension with the data driven approach\footnote{It is necessary to point out that the recent measurements of $e^+e^-\to \pi^+\pi^-$ cross section from threshold to 1.2 GeV with the CMD-3 detector \cite{cm3-230208834} will probably reduce the tension between lattice and data-driven approach. 
} \cite{Aoyama:2020ynm}.  
It is found that the light-quark connected part which gives about $90\%$ to HVP 
is a source for the tension. A lot of efforts by different lattice groups are made in order to investigate in detail different parts of this contribution to gain a clearer understanding of the difference \cite{Aubin:2022ino,Blum:2023qou,Wang:2022lkq,Ce:2022kxy}.  The whole story can be found in the ``White paper'' 2020 \cite{Aoyama:2020ynm} and updated in ``Snowmass 2021'' \cite{Colangelo:2022jxc}. 
 
The light-by-light (LbL) contribution via the non-perturbative QCD vacuum is suppressed in comparison to the HVP by the fine structure constant. Despite the smallness of this contribution, the theoretical understanding is important for an overall assessment of the full SM result. 

In the present paper, the LbL contribution of axial-vector particles is considered in the framework of the nonlocal quark model. 
The contribution of the axial vector meson to the muon anomalous magnetic moment is determined by the quark mechanism of a light meson production from a pair of photons. Due to the variety of quark models and nonperturbative effects therein, calculation of the above mentioned contribution remains an open question  \cite{Aoyama:2020ynm,Colangelo:2022jxc}.

The paper is a continuation of the quark model estimates of the LbL contributions.
In \cite{Dorokhov:2011zf,Dorokhov:2012qa,Dorokhov:2015psa}, the LbL contributions to the anomalous magnetic moment of the muon, the light pseudoscalar and scalar resonance exchange, and the quark loop within the nonlocal chiral quark model are calculated. 
Due to the nonlocality of interaction, the mass term in the quark propagator in the loops depends on the loop momentum.
Only ground states of mesons are considered in the model and the quark loop mimicks the excited states contribution.

In the present work, we generalize the calculation to include the vector--axial-vector sector\footnote{ 
It is interesting to note that in \cite{Dorokhov:2017nzk} it is shown that the axial-vector exchange interaction in muonic hydrogen makes an essential contribution to hyperfine splitting. The corrected hyperfine splitting is given in \cite{Miranda:2021lhb}. For hyperfine splitting, the axial-vector contribution is even larger than the pion contribution (see related results and uncertainties in \cite{Hagelstein:2015lph,Zhou:2015bea,Huong:2015naj,Dorokhov:2017gst}).
} 
refitting the model parameters to the observed pion data. The mixing of the pseudoscalar and longitudinal part of the axial-vector mesons as well as the $\rho-\gamma$ mixing are taken into account. Preliminary results are given in \cite{Dorokhov:2019tjc}.

The paper is structured as follows. In Sec.~\ref{model}, the nonlocal model is discussed.
Only non-strange mesons are considered. 
In Sec.~\ref{ExCurrents}, the interactions with external conserved currents are introduced. 
In Sec.~\ref{AVFF}, the two-photon transition form-factors of pseudoscalar and axial-vector mesons are considered.
Sec.~\ref{ModPatrams} is devoted to the discussion of model parameters.	
In Sec.~\ref{FormFactorParams}, the results for the two-photon form factor of the axial-vector meson are discussed.
In Sec.~\ref{lbl}, the procedure for calculating the light-by-light contribution from transition form-factors is presented.
The result of our calculations is given in Sec.~\ref{Result} and in Sec.~\ref{Discussion} with a comparison with other approaches.
The conclusions are given in Sec.~\ref{Conclusions}.	 

\section{Model}
\label{model}
The nonlocal chiral (light) quark model with the pseudoscalar--scalar and vector--axial-vector sectors is considered. The Lagrangian of the model has the form  
\begin{align}
&\mathcal{L}= \mathcal{L}_{free}+\mathcal{L}_{P,S}+\mathcal{L}_{V,A} 
 ,\quad 
\mathcal{L}_{free} = \bar{q}(x)(i \hat{\partial}-M_c)q(x), \\
&
\mathcal{L}_{P,S} = \frac{G_1}{2}\bigg(\Big(J_S^a(x)\Big)^2+\Big(J^a_P(x)\Big)^2\bigg) ,\quad 
\mathcal{L}_{V,A} = \frac{G_2}{2}\bigg(\Big(J_V^{a,\mu}(x)\Big)^2+\Big(J^{a,\mu}_{A}(x)\Big)^2\bigg),  \nonumber
\end{align}
where $M_c$ is the current quark mass matrix with diagonal elements $m_c$, $G_1$ and $G_2$ are the coupling constants in pseudoscalar--scalar (P,S) and vector--axial-vector sectors  (V,A), respectively. 

In the limit of vanishing current quark masses, the Lagrangian has a chiral symmetry similar to QCD.
The chiral symmetry is both spontaneously and explicitly broken by dynamical chiral symmetry breaking and nonzero current quark masses. As a result, in the mass spectrum there exist almost massless pseudo-Goldstone particles.

The nonlocal quark currents are given by \footnote{Such a structure of the interaction corresponds to that of the instanton liquid model \cite{Dorokhov:2000gu} (ILM).}
\begin{align}
J_{M}^{a\{,\mu\}}(x)=\int d^{4}x_{1}d^{4}x_{2}\,f(x_{1})f(x_{2})\, \bar{q}
(x-x_{1})\,\Gamma_{M}^{a\{,\mu\}}q(x+x_{2}),\label{eq2}
\end{align}
with $M=S,P,V,A$. The spin-flavour matrices are $\Gamma_{{S}}^{a}=\lambda^{a}$, $\Gamma_{{P}}^{a}=i\gamma^{5}\lambda^{a}$, $\Gamma_{{V}}^{a,\mu}=\gamma^{\mu}\lambda^{a}$, $\Gamma_{A}^{a,\mu}=\gamma^{5}\gamma^{\mu}\lambda^{a}$. For the $SU(2)$ model,  
the flavour matrices are: $\lambda^{a}\equiv\tau^{a}$, $a=0,..,3$ with $\tau^0=1$.
Such structure of interaction can be motivated by instanton liquid model \cite{Anikin:2000rq}.
$f(x)$ is the form factor encoding the nonlocality of the QCD vacuum. Since only four-quark interaction is considered, the action of the model can be bosonized by the usual Hubbard-Stratonovich trick with the introduction of auxiliary mesonic fields for each quark current, i.e. $P$, $S$, $V$, $A$.
The resulting effective Lagrangian after spontaneous symmetry breaking can be written in the form
\begin{align}
\mathcal{L}_{eff}&=
\bar{q}(x)(i \hat{\partial}_x -M_c)q(x)+\sigma_0 J_S^0(x) - \frac{1}{2 G_1}\left(\Big(P^a(x)\Big)^2+ \Big(\tilde{S}^a(x)+ \sigma_0 \delta^a_0\Big)^2\right)-\label{Bosonized}\\
&- \frac{1}{2 G_2}\left(\Big(V^{a,\mu}(x)\Big)^2 + \Big(A^{a,\mu}(x)\Big)^2\right)
+
P^a(x)J_{P}^{a}(x)+
\tilde{S}^{a}(x)J_{S}^{a}(x)+\notag\\
&+
V_\mu^a(x)J_{V}^{a,\mu}(x)+
A_\mu^a(x)J_{A}^{a,\mu}(x) .\notag
\end{align}
The scalar isoscalar field has a non-zero vacuum expectation value 
$\langle S^0 \rangle_0=\sigma_0\neq0$.
The shift of the scalar isoscalar field $S^0=\tilde{S}^0+\sigma^0$, which is necessary to obtain a physical scalar field with zero vacuum expectation value, leads to the appearance of the dynamical\footnote{Dynamical means that for small momentum the quark mass $m(0)$ is similar to the constituent one, while for large momentum it behaves like the current one $m_c$.
}
quark mass, which depends on the quark momentum\footnote{The same symbols are used for Fourier-transformed functions.} ($m_{d}=-\sigma^0$). 
The separable structure of quark current \eqref{eq2} leads to a solution where momentum dependence is factorized and "gap" equation takes the simple form
\begin{align}
m(p)=m_c+m_{d}f^2(p),\quad m_{d}= G_1 \frac{8  N_c }{(2 \pi)^4} \int d^4_Ek
\frac{f^2(k^2)m(k^2)}{k^2+m^2(k^2)}. \label{Gap}
\end{align}
This equation for scalar coefficient $m_{d}$ can be easily solved numerically.
The corresponding quark propagator is 
\begin{align}
\mathrm{S}(p)=(\hat{p}-m(p))^{-1}.
\end{align}

Meson propagators can be obtained by taking quadratic terms over the meson field from the Lagrangian at one loop level. For spin-0 mesons, the unrenormalized propagators are
\begin{align}
\mathrm{D}_M(p^2)=\frac{1}{-G_1^{-1}+\Pi_{MM}(p^{2})}= \frac{g^2_M(p^2)}{p^2-M_M^2}
	\label{pionpole},
\end{align}
the meson masses are located at points $p^2 = M_M^2$, which correspond to the solution of the equation
\begin{align}
	-G_1^{-1}+\Pi_{MM}\left(M_M^2\right)=0,\notag
\end{align}
and the value of the meson coupling constant on-mass shell $g_M(M_M^2)$ can be obtained from \eqref{pionpole} using l'H\^{o}pital's rule.  
After redefinition of the meson fields, the spin-0 propagator has the usual form 
\begin{align}
\mathrm{D}^R_M(p^2)=\mathrm{D}_M(p^2)/g^2_M(p^2)=(p^2-M_M^2)^{-1}. \label{DR}
\end{align}
The quark polarization loops are
\begin{align}
	&\Pi_{M_1M_2}(p^2)=i \frac{N_c}{(2\pi)^4} \int d^4 k
	f^2(k_+^2)f^2(k_-^2)\, \mathrm{Tr}_{d,f}\left[ \mathrm{S}(k_-)\Gamma_{M_1}^{a}
	\mathrm{S}(k_+) \Gamma_{M_2}^{b} \right], \notag 
\end{align}
where $ k _ \pm = k \pm p / 2$ and the trace is taken over Dirac and flavour matrices. This renormalization is important only for decay of mesons while for intermediate particles it is a completely identical procedure. 

Quark loops and propagators of vector and axial-vector mesons should be split into longitudinal and transverse parts
\begin{align}
\mathrm{D}^{\alpha\beta}_M(p^2)&=\mathrm{D}^{\mathrm{T}}_M(p^2)\mathrm{P}^{\mathrm{T};\alpha\beta}_{p}+\mathrm{D}^{\mathrm{L}}_M(p^2)\mathrm{P}^{\mathrm{L};\alpha\beta}_p,
\end{align}
with the help of appropriate projectors 
\begin{align}
\mathrm{P}^{\mathrm{T};\alpha\beta}_p &= g^{\alpha\beta}- \frac{p^\alpha p^\beta}{p^2},\,\, \mathrm{P}^{\mathrm{L};\alpha\beta}_p = \frac{p^\alpha p^\beta}{p^2}\notag.
\end{align}
Transverse components correspond to spin-1 states and unrenormalised propagators are
\begin{equation}
\mathrm{D}_{V,A}^{\mathrm{T}}(p^2)=\dfrac{1}{-G_{2}^{-1}+\Pi_{VV,AA}^{\mathrm{T}}(p^2)}= \frac{g^2_{V,A}(p^2)}{M_{V,A}^2-p^2},\quad 
\end{equation}
renormalized propagators are $\mathrm{D}^{\mathrm{T};R}_M(p^2)=\mathrm{D}^{\mathrm{T}}_M(p^2)/g^2_M(p^2)$ and the masses can be found from the solution of 
\begin{equation}
-G_{2}^{-1}+\Pi_{V,A}^{\text{T}}(M_{V,A}^2)=0	.\label{VMMassEq}%
\end{equation}

On the other hand, the vertex functions and the meson masses can be found
from the Bethe–Salpeter equation, which for the pion case is
\begin{equation}
\delta(p_1 + p_2 - p_3 - p_4)\frac{\overline{\mathbf{\Gamma}}^{\pi}_{p_1,p_3}\otimes
		\mathbf{\Gamma}^{\pi}_{p_2,p_4}}{p^{2}-\mathrm{M}_{{\pi}}^{2}},
\end{equation}%
where 
$p$ is the total momentum of the $\bar{q}q$ pair, $\overline{\mathbf{\Gamma}}=\gamma^0\mathbf{\Gamma}^\dag \gamma^0$ and $p_i$ are quark momenta.
The meson vertex functions without mixing in momentum space are
\begin{align}
	\mathbf{\Gamma}^{M\{;\mu\}}_{p_+,p_-}=g_{M}(p^2) \Gamma_M^{\{\mu\}} f(p_-) f(p_+), \label{GMesonQQ}
\end{align}
where 
$p_\pm,k$ are the quark and meson momenta, respectively. 

Longitudinal components are related to spin-0.
In the case of a system of pseudoscalar--axial-vector states, a mixing \cite{Volkov:1986zb,Meissner:1987ge} appears due to a quark polarisation loop with pseudoscalar and axial-vector vertices 
\begin{align}
	\Pi^\mu_{PA}(p^2)=p^\mu \Pi_{\pi a_1}(p^2),
\end{align}
and physical states can be found as solutions of the matrix equation \cite{Kaloshin:2001sn}:
\begin{align}
\mathrm{\tilde{D}}_{P}(p^2)&=\frac{-G_2^{-1}+\Pi^L_{AA}(p^2)}{\mathcal{D}(p^2)},\,
\mathrm{\tilde{D}}_{PA}(p^2)=\frac{\Pi_{PA}(p^2)}{\mathcal{D}(p^2)},\,
\mathrm{\tilde{D}}^L_{A}(p^2)=\frac{-G_1^{-1}+\Pi_{PP}(p^2)}{\mathcal{D}(p^2)},\label{PA}\\
\mathcal{D}(p^2)&=\left[-G_1^{-1}+\Pi_{PP}(p^2)\right]\left[-G_2^{-1}+\Pi^L_{AA}(p^2)\right]-p^2 \Pi^2_{PA}(p^2), \quad  \mathrm{\tilde{D}}^\mu_{PA}(p^2)=p^\mu \mathrm{\tilde{D}}_{\pi a_1}(p^2). \notag
\end{align}

One can represent the mixing \eqref{PA} as a modification of the pion vertex with the contribution of the longitudinal component of the axial-vector mesons
\begin{align}
\mathbf{\Gamma}^\pi_{p_+,p_-}= i\gamma^{5}\lambda^{a}\left(g_{\pi}(p^2)-\hat{p}\tilde{g}_{\pi}(p^2)\right) f(p_-) f(p_+). \label{PionModifiedVertex}
\end{align}
The pion coupling constants can be obtained by comparing the $T$-matrix elements from \eqref{PionModifiedVertex} and the solution of the system \eqref{PA}
\begin{align}
\frac{g^2_{\pi}(p^2)}{p^2-M_\pi^2}=\mathrm{\tilde{D}}_{P}(p^2), \quad \frac{g_{\pi}(p^2)\tilde{g}_{\pi}(p^2)}{p^2-M_\pi^2}=\mathrm{\tilde{D}}_{PA}(p^2).
\end{align}
Then the rest of the contribution of the longitudinal part of the axial-vector meson is simply
\begin{align}
\mathrm{\tilde{D}}^L_{A}(p^2)-p^2\frac{\tilde{g}_{\pi}^2(p^2)}{p^2-M_\pi^2}=\frac{1}{-G_2^{-1}+\Pi^L_{AA}(p^2)}=\mathrm{{D}}^L_{A}(p^2),
\end{align}
i.e. all mixing is ``eaten'' by the modification of the pion vertex. 

Since the pion is a Goldstone boson, it should be massless in the case of exact chirally symmetry, i.e. by setting $m_c$ to zero. In nonlocal model with only scalar--pseudoscalar sector, it is shown how one can reproduce the Gell-Mann–Oakes–Renner relation
\begin{align}
M_\pi^2 f_\pi^2 = -2m_c\langle \bar{q}q \rangle
\end{align}
analytically with help of chiral expansion \cite{Osipov:2007zz}. The mixing with axial-vector meson does not change the Goldstone nature of pion \cite{Plant:1997jr}.
A similar detailed derivation on the basis of chiral expansion for the nonlocal model with mixing will be presented elsewhere.

\section{External currents}
\label{ExCurrents}
Due to nonlocality, the interactions with the electromagnetic gauge field should be introduced not only in the quark kinetic part but also in the nonlocal quark currents. Thus, in the presence of external gauge fields, the part of Lagrangian for meson fields with quark currents in Eq. (\ref{eq2}) takes the form 
\begin{align}
M^{a\{,\mu\}}(x) J_{M}^{a\{,\mu\}}(x)&=\int d^4x_1 d^4x_2 \, f(x_1)f(x_2)\, \bar{Q}(x-x_1,x) \, M^{a\{,\mu\}}(x)\Gamma_M^{a\{,\mu\}}  \, Q(x,x+x_2) , \label{JwithQ} 
\end{align}
where the Schwinger phase factor\footnote{In case of non-Abelian external field, the exponent should be path-ordered.} is attached to each quark field $Q(x,x+x_2)= E(x,x+x_2)q(x+x_2)$ and $\bar{Q}(x-x_1,x)= \bar{q}(x-x_1)E(x-x_1,x)$
\begin{align}
	&E(x,y)=\mathrm{exp}\left\{-i\mathrm{e}\mathrm{Q}\int \limits_x^y du_\mu G^\mu(u)\right\}, \label{Pexp}
\end{align}
where $G^\mu$ is the photon field, $\mathrm{e}$ is the elementary charge and $\mathrm{Q}$ is the charge matrix of the quark fields.
The Eq. \eqref{JwithQ} is invariant under a gauge transformation 
\begin{align}
q(x) &= \mathrm{exp}\left\{i \alpha(x)\mathrm{Q}\right\}q'(x) ,\notag\\
\bar{q}(x) &= \bar{q}'(x)\mathrm{exp}\left\{-i \alpha(x)\mathrm{Q}\right\} ,\\
G_\mu(x) &= G'_\mu(x)+\frac{1}{\mathrm{e}} \partial_\mu\alpha(x)\notag ,\\
M^{a\{,\mu\}}(x)\lambda^{a} &=\mathrm{exp}\left\{i \alpha(x)\mathrm{Q}\right\} M'^{a\{,\mu\}}(x)\lambda^{a}\left\{-i \alpha(x)\mathrm{Q}\right\},\notag
\end{align}
since the Schwinger phase factor is transformed as 
\begin{align}
E(x,y)&=\mathrm{exp}\left\{+i\alpha(x)\mathrm{Q}\right\}\mathrm{exp}\left\{-i\mathrm{e}\mathrm{Q}\int \limits_x^y du_\mu G'^\mu(u)\right\}\mathrm{exp}\left\{-i\alpha(y)\mathrm{Q}\right\}=\notag\\
&=\mathrm{exp}\left\{+i\alpha(x)\mathrm{Q}\right\} E'(x,y)\mathrm{exp}\left\{-i\alpha(y)\mathrm{Q}\right\}
. \label{PexpGauge}
\end{align}
The gauge invariance leads to the Ward identity. 
Unfortunately, the Ward identity only fixes the longitudinal part of photon vertices, and to find an expression for the transverse part of vertices one needs to specify rules for the contour integral. One of the possible ways is to use the straight-path ansatz   \cite{Bowler:1994ir,Plant:1997jr,Scarpettini:2003fj}  
$z^{\mu}=x^{\mu}+\alpha(y^{\mu}-x^{\mu})$, $0\leq\alpha\leq 1$. 
An alternative scheme \cite{Terning:1991yt}, which is used in this paper, is based on the rules according to which the derivative of the contour integral does not depend on the form of the path and the explicit form of the path is not important 
\begin{align}
	\frac{\partial}{\partial y^{\mu }}\int\limits_{x}^{y}dz_\nu G^\nu(z)=G_{\mu }(y),\quad \delta^{(4)}\left( x-y\right)
	\int\limits_{x}^{y}dz_\nu G^{\nu }(z)=0.
\end{align}
The crucial feature of such prescriptions is that the resulting expression for diagrams with nonlocal vertices is expressed by finite-differences with momenta of diagrams with local vertices.

\begin{figure}[t]
\centering
\begin{center}
\begin{tabular*}{0.5\textwidth}{@{}ccccc@{}}
	\raisebox{-0.5\height}{\resizebox{0.15\textwidth}{!}{\includegraphics{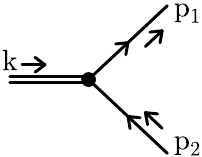}}}&{ }  { } { }  { }&
	\raisebox{-0.5\height}{\resizebox{0.15\textwidth}{!}{\includegraphics{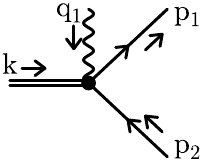}}}&{ }  { } { } { }  &
	\raisebox{-0.5\height}{\resizebox{0.15\textwidth}{!}{\includegraphics{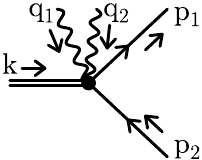}}}\\
	$(a)$&&$(b)$&&$(c)$
\end{tabular*}
\end{center}	
\caption{Meson--quark--antiquark vertices: without photon $(a)$, with one photon $(b)$ and with two photons $(c)$.}
\label{fig:MesonV}
\end{figure}

The Schwinger phase gauge factor contains the field in the exponent. Thus, vertices with arbitrary number of photon fields can be generated. Due to the nonlocal interaction between mesons and quarks, Fig. \ref{fig:MesonV}a, the vertices with one or two photons, Fig. \ref{fig:MesonV}bc, take the form ($p_1=p_2+q_1+...+q_i$)
\begin{align}
&\mathbf{\Gamma}^{M\gamma ;\{\alpha\}\mu}_{p_1,p_2,q_1}= -\mathrm{e}g_{M}(k)\bigg(J_1^\mu(p_1,-q_1)
\mathrm{Q}\Gamma_{M}^{a\{,\alpha\}} f(p_2)
+f(p_1)\Gamma_{M}^{a\{,\alpha\}}\mathrm{Q} J_1^\mu(p_2,q_1)\bigg),\notag\\
&\mathbf{\Gamma}^{M\gamma\gamma;\{\alpha\}\mu\nu}_{p_1,p_2,q_1,q_2}=\mathrm{e}^2g_{M}(k)\bigg(J_2^{\mu\nu}(p_1,-q_1,-q_2) \mathrm{Q}^2\Gamma_{M}^{a\{,\alpha\}} f(p_2)
+ J_1^\mu(p_1,-q_1)\mathrm{Q}\Gamma_{M}^{a\{,\alpha\}}\mathrm{Q}J_1^\nu(p_2,q_2)+\notag\\
&\quad\quad
+J_1^\nu(p_1,-q_2)\mathrm{Q}\Gamma_{M}^{a\{,\alpha\}}\mathrm{Q}J_1^\mu(p_2,q_1)
+ f(p_1)\Gamma_{M}^{a\{,\alpha\}}\mathrm{Q}^2J_2^{\mu\nu}(p_2,q_1,q_2)\bigg),\notag\\
&\quad 
J_1^\mu(p,q) = (2p+q)^\mu\mathrm{f}^{(1)}_{p+q,p},\notag\\
&\quad
J_2^{\mu\nu}(p,q_1,q_2) =+2g^{\mu\nu} \mathrm{f}^{(1)}_{p,p+q_1+q_2}+(2p+q_1)^\mu(2p +2q_1+q_2)^\nu\mathrm{f}^{(2)}_{p,p+q_1,p+q_1+q_2}+\notag\\
&\quad\quad
+(2p+q_2)^\nu(2p +2q_2+q_1)^\nu\mathrm{f}^{(2)}_{p,p+q_2,p+q_1+q_2}, \label{MesonGG}
\end{align}
where the shorthand notations for first and second order finite-differences are introduced\footnote{ 
Similar abbreviations are used for finite differences of the mass function $m\rightarrow \mathrm{m}^{(i)}$
.}
\begin{align}
&\mathrm{f}^{(1)}_{p,q}     =\frac{f\left(  p\right)
		-f\left(  q\right)  }{p^{2}-q^{2}},\quad
\mathrm{f}^{(2)}_{p,q,l}     =\frac{\mathrm{f}^{(1)}_{p,q}  -\mathrm{f}^{(1)}_{p,l}}{q^2-l^2}.\nonumber
\end{align}

The vacuum expectation terms of the scalar field generate antiquark-quark-photon(s) vertices, which can be rewritten from the expression for the scalar current \eqref{MesonGG} by using Eq. \eqref{Gap} in terms of the quark mass
\begin{align}
&{\Gamma}^{\gamma;\mu}_{p_1,p_2,q}=\mathrm{e}\mathrm{Q} \bigg(\gamma_\mu-(p_{1}+p_{2})_{\mu}\mathrm{m}^{(1)}_{p_1,p_2}\bigg), \nonumber\\
&{\Gamma}^{\gamma\gamma;\mu\nu}_{p_1,p_2,q_1,q_2}=\mathrm{e}^2 \mathrm{Q}^2\bigg(
 2g^{\mu\nu}\mathrm{m}^{(1)}_{p_1,p_2}
+ (2p_2+q_1)^\mu (2p_1-q_2)^\nu\mathrm{m}^{(2)}_{p_1,p_1+q_1,p_2}\notag \\
&\quad \qquad\qquad\qquad\qquad\qquad\quad+ (2p_2+q_2)^\nu (2p_1-q_1)^\mu\mathrm{m}^{(2)}_{p_1,p_1+q_2,p_2}
\bigg).
\end{align}

\begin{figure}[t]
\centering
\begin{center}
\begin{tabular*}{0.65\textwidth}{@{}ccccccc@{}}
\raisebox{-0.5\height}{\resizebox{0.185\textwidth}{!}{\includegraphics{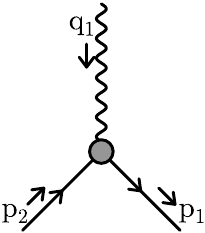}}}&{ }={ } &
\raisebox{-0.5\height}{\resizebox{0.15\textwidth}{!}{\includegraphics{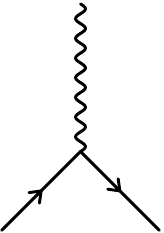}}}&{ }+{ } &
\raisebox{-0.5\height}{\resizebox{0.15\textwidth}{!}{\includegraphics{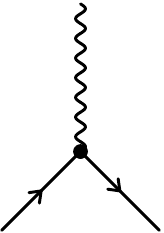}}}&{ }+{ } &
\raisebox{-0.5\height}{\resizebox{0.15\textwidth}{!}{\includegraphics{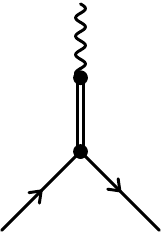}}}\\
			$(a)$&&$(b)$&&$(c)$&&$(d)$
\end{tabular*}
\end{center}	
\caption{Quark--antiquark--photon vertices : full $(a)$, with local vertex $(b)$, with nonlocal vertex $(c)$  and with vector meson--photon transition $(d)$.}
\label{fig:GGaqq}
\end{figure}

\begin{figure}[t]
\centering
\begin{center}
\begin{tabular*}{0.65\textwidth}{@{}ccccccc@{}}
	\raisebox{-0.5\height}{\resizebox{0.15\textwidth}{!}{\includegraphics{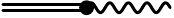}}}&{ }={ } &
	\raisebox{-0.5\height}{\resizebox{0.2\textwidth}{!}{\includegraphics{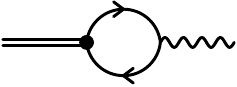}}}&{ }+{ } &
	\raisebox{-0.5\height}{\resizebox{0.2\textwidth}{!}{\includegraphics{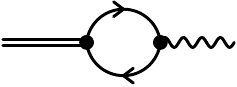}}}&{ }+{ } &
	\raisebox{-0.5\height}{\resizebox{0.15\textwidth}{!}{\includegraphics{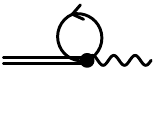}}}\\
		$(a)$&&$(b)$&&$(c)$&&$(d)$
\end{tabular*}
\end{center}	
\caption{Vector meson--photon mixing diagrams: total $(a)$, with local vertex $(b)$ and with nonlocal vertices $(c)$, $(d)$.}
\label{fig:RhoGamma}
\end{figure}

In the presence of the vector sector, the photon(s)-quark interaction vertices are additionally dressed by the  $\rho(\omega)\rightarrow \gamma$ transition \cite{Plant:1997jr,Dorokhov:2003kf}, see Fig.\ref{fig:RhoGamma}. This dressing is transversal and can be written in the form
\begin{align}
\label{PROV}
&C_{\gamma V}(q^2)=  i
N_{\text{c}} \frac{\mathrm{P}^{T;\mu\nu}_{q}}{3}\int\dfrac{\mathrm{d}^{4}k}{(2\pi)^{4}}\left\{  \mathrm{Tr}%
\left[  \mathrm{S}(k_{+})\mathbf{\Gamma}^{\gamma;\mu}_{k_{+},k_{-},q} \mathrm{S}(k_{-})	\mathbf{\Gamma}^{V;\nu}_{k_-,k_+}\right]
+
 \mathrm{Tr}\left[ \mathbf{\Gamma}^{M\gamma ;\mu\nu}_{k,k,q} \mathrm{S}(k)\right]  \right\}  ,
\end{align}
where $V$ stands for $\rho^{0}$ or $\omega$ mesons. The transition has the property $C_{\gamma V}(0)=0$ \cite{Plant:1997jr} and does not lead to the renormalization of photon mass or quark charge \cite{Dorokhov:2003kf}.

One can understand the effect of dressing at the diagram level by joining Fig. \ref{fig:RhoGamma} with  Fig. \ref{fig:MesonV}a or  Fig. \ref{fig:MesonV}b to get the full expression for vertices with one or two photons 
\begin{align}
&\mathbf{\Gamma}^{\gamma;\mu}_{p_2,p_1,q}={\Gamma}^{\gamma;\mu}_{p_2,p_1,q} + \sum\limits_{V=\rho^0,\omega}\mathbf{\Gamma}^{V;\alpha}_{p_2,p_1}\mathrm{P}^{\mathrm{T};\alpha\mu}_{q} C_{\gamma V}(q), \notag\\
&\mathbf{\Gamma}^{\gamma\gamma;\mu\nu}_{p_2,p_1,q_1,q_2}={\Gamma}^{\gamma\gamma;\mu\nu}_{p_2,p_1,q_1,q_2} + \sum\limits_{V=\rho^0,\omega}\mathbf{\Gamma}^{V\gamma ;\alpha\nu}_{p_2,p_1,q_1}\mathrm{P}^{\mathrm{T};\alpha\mu}_{q_1}  C_{\gamma V}(q_1) + \sum\limits_{V=\rho^0,\omega}\mathbf{\Gamma}^{V\gamma ;\alpha\mu}_{p_2,p_1,q_2}\mathrm{P}^{\mathrm{T};\alpha\nu}_{q_2} C_{\gamma V}(q_2).
\end{align}

It is important to note that from Eq. \eqref{JwithQ}, only one vector meson can be connected with quark-antiquark pair and external EM fields at the point of interaction, as shown in Fig. \ref{fig:GGaqq}(c,d). This fact is also critical for describing the two-photon vertex of interaction with quark fields, taking into account the $\rho-\gamma$ mixing.

\begin{figure}[t]
\centering
\begin{center}
\begin{tabular*}{0.65\textwidth}{@{}ccccccc@{}}
	\raisebox{-0.5\height}{\resizebox{0.185\textwidth}{!}{\includegraphics{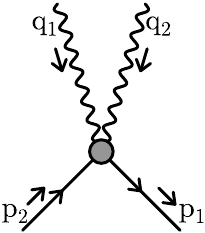}}}&{ }={ } &
	\raisebox{-0.5\height}{\resizebox{0.15\textwidth}{!}{\includegraphics{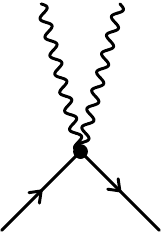}}}&{ }+{ } &
	\raisebox{-0.5\height}{\resizebox{0.15\textwidth}{!}{\includegraphics{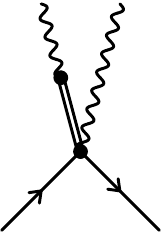}}}&{ }+{ } &
	\raisebox{-0.5\height}{\resizebox{0.15\textwidth}{!}{\includegraphics{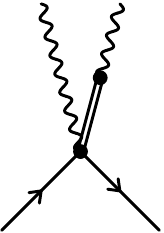}}}\\
	$(a)$&&$(b)$&&$(c)$&&$(d)$
\end{tabular*}
\end{center}	
\caption{Quark--antiquark--photon--photon vertices:  with nonlocal vertex $(b)$ and with vector meson--photon transition $(c)$, $(d)$.}
\label{fig:GGaGaqq}
\end{figure}

\begin{figure}[t]
\centering
\begin{center}
	\begin{tabular*}{1\textwidth}{@{}ccccccccc@{}}
		\raisebox{-0.5\height}{\resizebox{0.17\textwidth}{!}{\includegraphics{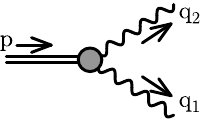}}}&{ }={ } &
		\raisebox{-0.5\height}{\resizebox{0.2\textwidth}{!}{\includegraphics{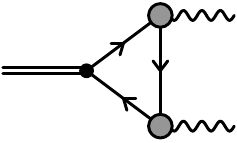}}}&{ }+{ } &
		\raisebox{-0.5\height}{\resizebox{0.2\textwidth}{!}{\includegraphics{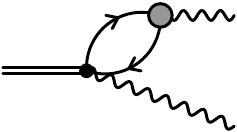}}}&{ }+{ } &
		\raisebox{-0.5\height}{\resizebox{0.2\textwidth}{!}{\includegraphics{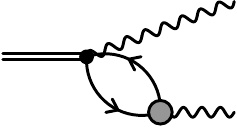}}}&{ }+{ } &\\
		$(a)$&&$(b)$&&$(c)$&&$(d)$\\
&{ } { }&{ }  { }&{ }+{ } &\raisebox{-0.5\height}{\resizebox{0.2\textwidth}{!}{\includegraphics{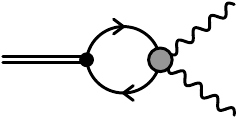}}}
&{ }+{ } &\raisebox{-0.5\height}{\resizebox{0.15\textwidth}{!}{\includegraphics{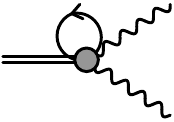}}}
\\	
		&& &&$(e)$&&$(f)$\\
\end{tabular*}
\end{center}	
\caption{Diagrams for meson transition form-factor: quark triangle $(b)$ and with meson--photon--quark--antiquark vertex transition $(c)$, $(d)$, quark--antiquark--two--photon $(e)$ and meson--quark--antiquark--two--photon $(f)$.}
\label{fig:GaMGaGa}
\end{figure}

\section{AV form-factor}
\label{AVFF}

The transition form-factor of the pseudoscalar meson has only one structure
\begin{align}
&T^{\mu\nu}\left( p, q_{1}, q_{2}\right)=e^{2}\Delta_P^{\mu\nu}\left( p, q_{1}, q_{2}\right), \quad  \Delta_P^{\mu\nu}\left( p, q_{1}, q_{2}\right)=\varepsilon_{\mu\nu\rho\sigma}q_{1}^{\rho}q_{2}^{\sigma} \mathrm{F}_{P}\left(  p^{2};q_{1}^{2},q_{2}^{2}\right) ,
\end{align}
where $p$ is the momentum of virtual meson and the two photons have momenta $q_{1,2}$.

The general expression of the axial-vector meson transition form factor is \cite{Rosenberg:1962pp,Adler:1969gk}
\begin{align}
&T^{\mu\nu}_\alpha\left( p, q_{1}, q_{2}\right)=e^{2}\Delta_{A,\alpha}^{\mu\nu}\left( p, q_{1}, q_{2}\right),\quad  
\Delta_{A,\alpha}^{\mu\nu}\left( p, q_{1}, q_{2}\right) =i\varepsilon_{\rho \sigma \tau \alpha} \sum_{i=1}^{6}A_i(p^2,q_1^2,q_2^2)B_i^{\mu\nu \rho\sigma\tau}\notag \\ 
&\quad\quad 
B_1 = q_1^{\tau}g^{\mu \rho}g^{\sigma \nu},\quad 
B_2 = q_2^{\tau}g^{\mu \rho}g^{\sigma \nu},\quad 
B_3 = q_1^{\nu} q_1^{\rho} q_2^{\sigma}g^{\tau \mu} \nonumber\\
&\quad\quad
B_4 = q_2^{\nu} q_1^{\rho} q_2^{\sigma}g^{\tau \mu},\quad 
B_5 = q_1^{\mu} q_1^{\rho} q_2^{\sigma}g^{\tau \nu},\quad 
B_6 = q_2^{\mu} q_1^{\rho} q_2^{\sigma}g^{\tau \nu} \label{RosenbergAdler}
\end{align}
where $p$, $q_1$ and $q_2$ are the momenta of the AV meson and photons with indices $\alpha,\mu,\nu$.
Gauge invariance leads to the relations
\begin{align}
A_2(p^2,q_1^2,q_2^2) &=  (q_1\cdot q_2)A_6(p^2,q_1^2,q_2^2)+q_1^2A_5(p^2,q_1^2,q_2^2) , \notag \\
A_1(p^2,q_1^2,q_2^2) &= (q_1\cdot q_2)A_3(p^2,q_1^2,q_2^2) + q_2^2 A_4(p^2,q_1^2,q_2^2), \label{AVgauge}
\end{align}
and the Bose symmetry results in:
\begin{align}
{A}_1(p^2,q_1^2,q_2^2)&=-{A}_2(p^2,q_2^2,q_1^2),  \quad 
{A}_3(p^2,q_1^2,q_2^2)=-{A}_6(p^2,q_2^2,q_1^2), \notag \\
{A}_4(p^2,q_1^2,q_2^2)&=-{A}_5(p^2,q_2^2,q_1^2). 
\end{align}
The part of the amplitude longitudinal to the meson momentum is
\begin{align}
\Delta_{A,\alpha;L}^{\mu\nu}\left( p, q_{1}, q_{2}\right)=	i {\varepsilon^{\rho\sigma\mu\nu}}q_{1{\rho}} q_{2\sigma}p_\alpha \frac{1}{p^2} \bigg({A}_2(p^2,q_1^2,q_2^2)-{A}_1(p^2,q_1^2,q_2^2)\bigg).
\end{align}
The transverse part of the amplitude can be rewritten as\footnote{
Various expressions for the transition form factor can be found in \cite{Cahn:1986qg,Rudenko2017,Osipov:2017ray,Szczurek:2020hpc,Hoferichter:2020lap,Pascalutsa:2012pr,Zanke:2021wiq}.
} \cite{Pascalutsa:2012pr}:
\begin{align}
&\Delta_{A,\alpha}^{\mu\nu}\left( p, q_{1}, q_{2}\right) =i 
	\varepsilon_{\rho \sigma \tau \alpha}
\biggl\{
R^{\mu \rho}_{ q_1, q_2} R^{\nu \sigma}_{ q_1, q_2} \,
(q_1 - q_2)^\tau \, {(q_1 \cdot q_2)} \, F^{(0)}_{A\gamma^\ast\gamma^\ast}(p^2,q_1^2, q_2^2)
\nonumber \\
&\quad\quad\quad + \, R^{\nu \rho}_{ q_1, q_2} Q_1^\mu
q_1^\sigma \, q_2^\tau \,   \, F_{A\gamma^\ast\gamma^\ast}^{(1)}(p^2,q_1^2, q_2^2) 
 +R^{\mu \rho}_{ q_1, q_2} Q_2^\nu
q_2^\sigma \, q_1^\tau \,  \, F^{(1)}_{A\gamma^\ast\gamma^\ast}(p^2,q_2^2, q_1^2) \biggr\}
, \nonumber \\
&\quad R^{\mu \nu} _{ q_1, q_2}  = - g^{\mu \nu} + \frac{1}{X} \,
\bigl \{ (q_1 \cdot q_2)\left( q_1^\mu \, q_2^\nu + q_2^\mu \, q_1^\nu \right)
- q_1^2 \, q_2^\mu \, q_2^\nu  - q_2^2 \, q_1^\mu \, q_1^\nu \bigr\}
,  \label{Pascalutsa}\\
&\quad \quad Q_1^\mu=
q_1^\mu -q_2^{\mu}  \frac{q_1^2}{(q_1 \cdot q_2)} 
,\quad
Q_2^\nu=
q_2^\nu - q_1^{\nu} \frac{q_2^2}{(q_1 \cdot q_2)}
,\quad X = (q_1 \cdot q_2)^2 - q_1^2 q_2^2,\nonumber
\end{align}
where $R^{\mu \nu} _{ q_1, q_2}$  is the totally transverse tensor,  $Q_1^\mu$ and $Q_2^\nu$ are transverse with respect to $q_1$ and $q_2$, respectively\footnote{Our definition differs from  \cite{Pascalutsa:2012pr} by a factor $M_A^2$.}. 
After projecting to the transverse components and using the Shouten identity, one can relate \eqref{RosenbergAdler} and \eqref{Pascalutsa} as
\begin{align}
F^{(0)}_{A\gamma^\ast\gamma^\ast} &=\frac{(q_1^2 + (q_1 \cdot q_2)) A_1+
((q_1 \cdot q_2) + q_2^2)(q_1^2A_5 + (q_1 \cdot q_2)A_6 ) }{(q_1 \cdot q_2) (q_1^2 - q_2^2)}
, \label{eq-F0F1} \\
F^{(1)}_{A\gamma^\ast\gamma^\ast} &
=-\frac{(q_1 \cdot q_2)}{X}\bigg( A_1  + (q_1 \cdot q_2)A_5 
+ q_2^2 A_6\bigg).\notag
\end{align}
The other set of Lorentz structures is suggested in \cite{Hoferichter:2020lap}, where the asymptotic behavior of meson transition form factors from a light-cone expansion is studied. The relations with scalar functions $\mathcal{F}_i^A$ from \cite{Hoferichter:2020lap} are:
\begin{align}
\mathcal{F}_1^A = M_A^2(A_3 + A_6)/2,\, \mathcal{F}_2^A = -M_A^2(A_3 + A_5),\, \mathcal{F}_3^A = -M_A^2(A_4 + A_6). \label{eqF2}
\end{align}
In \eqref{eq-F0F1} and \eqref{eqF2} the arguments of functions $F^{(i)}_{A\gamma^\ast\gamma^\ast}$, $\mathcal{F}_i^A$ and $A_i$ are $(p^2,q_1^2, q_2^2)$. 

According to the Landau--Yang theorem \cite{Landau:1948kw,Yang:1950rg}, the axial-vector mesons cannot decay into two real photons. However, the coupling of $1^{++}$ mesons to two photons is allowed if one or both photons are virtual. The two-photon ``decay'' width for axial-vector mesons is defined for a quasireal longitudinal photon and a real photon as
\begin{align}
&\tilde\Gamma_{\gamma\gamma}(A) = \lim_{Q^2\rightarrow 0} \frac{1}{2}\frac{M_A^2}{Q^2} \Gamma\left(A\rightarrow \gamma_T\gamma_L^\ast\right) =\frac{\pi\alpha^2M_{A}^5}{12}[F^{(1)}_{A\gamma^\ast\gamma^\ast}(M_{A}^2,0,0)]^2. 
\label{GammaWidth}
\end{align}

In the quark model, the photon-meson transition amplitude is a sum of the diagrams shown in Fig. \ref{fig:GaMGaGa}. The general expression for the quark loop integral has the form
\begin{align}
\Delta&_{M,\{\alpha\}}^{\mu\nu}\left( p, q_{1}, q_{2}\right)= -i N_{c}
\int\frac{d^{4}k}{(2\pi)^{4}}  
\mathrm{Tr} \bigg(2 \mathbf{\Gamma}^{M;\{\alpha\}}_{k_1,k_2}\mathrm{S}(k_1)
\mathbf{\Gamma}^{\gamma;\mu}_{k_1,k_3,-q_1}\mathrm{S}(k_3)
\mathbf{\Gamma}^{\gamma;\nu}_{k_3,k_2,-q_2} \mathrm{S}(k_2)+\nonumber\\
&\quad+ 
\mathbf{\Gamma}^{M\gamma ;\{\alpha\}\mu}_{k_2,k_3,-q_1}\mathrm{S}(k_3)\mathbf{\Gamma}^{\gamma;\nu}_{k_3,k_2,-q_2}\mathrm{S}(k_2)
+ 
\mathbf{\Gamma}^{M\gamma ;\{\alpha\}\nu}_{k_3,k_1,-q_2}S(k_1)\mathbf{\Gamma}^{\gamma;\mu}_{k_1,k_3,-q_1}\mathrm{S}(k_3) \nonumber\\
&\quad+ \mathbf{\Gamma}^{M;\{\alpha\}}_{k_1,k_2}\mathrm{S}(k_1)\mathbf{\Gamma}^{\gamma\gamma;\mu\nu}_{k_1,k_2,-q_1,-q_2}\mathrm{S}(k_2) + \mathbf{\Gamma}^{M\gamma\gamma;\{\alpha\}\mu\nu}_{k_3,k_3,-q_1,-q_2}\mathrm{S}(k_3) \bigg),\label{DeltaVertex}
\end{align}
where the symbols are the photon momenta $q_{1,2}$, the meson momentum $p=q_{1}+q_{2}$, and the quark momenta $k_{1,2,3}$ ($k_{1}=k+q_{1}$, $k_{2}=k-q_{2}$, $k_3=k$). The first term in parentheses corresponds to the quark triangle diagram in Fig. \ref{fig:GaMGaGa}b (factor 2 comes from crossed one) and the next terms correspond to the diagrams in Figs. \ref{fig:GaMGaGa}c-f with effective nonlocal vertices. 

\section{Model Parameters}
\label{ModPatrams}	
At first sight, the model parameters can be taken from \cite{Plant:1997jr}, since our model is somewhat similar: only non-strange particles with vector–axial-vector sector, form-factor in Gaussian form, $f(k^2)=\exp(-k^2/\Lambda^2)$ in Euclidean space, and currents with derivatives which are leading to the quark wave function renormalization are not included as in \cite{IzzoVillafane:2016mhc,Carlomagno:2019yvi}. 

\begin{table*}
\caption{
Different parameterizations of the model parameters $\Lambda$, $m_c$, $m_d$, dimensionless combinations $G_1{\Lambda^2}$, $G_2{\Lambda^2}$ and position of the first two poles of quark propagator in complex plane. For comparison, the model parameters for the model without vector particles using the same values of $m_d$ are also presented and denoted by asterisk.}
\begin{tabular}{|ccccccc||ccc|}
		\hline
		$     m_{d}$, & $     m_{c}$, & $    \Lambda $, & 		\multirow{2}{*}{$  G_1{\Lambda^2}$} &  		\multirow{2}{*}{$  G_2{\Lambda^2}$} & $m^2_{\mathrm{pole}}$, & $\mathrm{M}_{\mathrm{thr}}$,& $ m^*_{c}$, & $  \Lambda^*  $, &\multirow{2}{*}{$    G^*_1{{\Lambda^*}^2}$}    \\
		MeV &      MeV & MeV  &  & & GeV$^{2}$  &  MeV &  MeV & MeV &	\\
		\hline
		$           293$ & $            7.12$ & $            1066.5$ & $            34.808$ & $            -3.00$  & $0.152$; $0.489$         & $780$ &   $ 7.64$          & $           1029.3$ & $            35.144$     \\
		$           300$ & $            7.28$ & $            1045.5$ & $            35.357$ & $            -4.36$  & $0.180$; $0.404$         & $848$ &  $ 8.09$          & $            992.2$ & $            35.888$     \\
		$           310$ & $            7.72$ & $            1004.8$ & $            36.297$ & $            -5.02$  & $0.245 \pm 0.091$        & $1006$& $ 8.73$          & $            945.3$ & $            36.980$     \\
		$           320$ & $            8.24$ & $             963.1$ & $            37.333$ & $            -5.20$  & $0.202 \pm 0.152$        & $954$ & $ 9.37$          & $            903.8$ & $            38.115$     \\
		$           330$ & $            8.82$ & $             922.3$ & $            38.453$ & $            -5.09$  & $0.164 \pm 0.180$        & $902$ & $10.01$          & $            866.6$ & $            39.293$     \\
		$           340$ & $            9.48$ & $             882.5$ & $            39.665$ & $            -4.72$  & $0.131 \pm 0.194$        & $854$ &$10.65$          & $            833.0$ & $            40.515$     \\
		$           350$ & $           10.24$ & $             842.5$ & $            40.999$ & $            -3.98$  & $0.101 \pm 0.200$        & $806$ &$11.29$          & $            802.6$ & $            41.776$     \\
		$           354$ & $           10.64$ & $             824.2$ & $            41.615$ & $            -3.33$  & $0.088 \pm 0.201$        & $784$ &$11.54$          & $            791.5$ & $            42.286$     \\
		\hline
	\end{tabular}
	\label{FitModParams}
\end{table*}

However, it is better to study the dependence of LbL contribution similar to the analysis in  \cite{Dorokhov:2012qa,Dorokhov:2015psa}. Recall that the scalar-pseudoscalar sector model has three parameters: the current quark mass $m_{c}$, the dynamical quark mass $m_{d}$ and the nonlocality parameter $\Lambda$. 

In principle one can use the PDG value \cite{ParticleDataGroup:2022pth} to fix the current quark mass.
However it is not a simple procedure since the actual scale of the model is unknown and therefore it is not clear how to evolve the values from the scale of $2$ GeV. It is expected that the model scale should be of the order of $1$ GeV, however even smaller values for model scale are discussed in the literature \cite{Broniowski:2007si}.
Due to this complication the current quark mass is considered as an independent model parameter.

To understand the stability of the model predictions with respect to changes in the model parameters, one can vary the dynamical quark mass in a rather wide physically acceptable interval of $200$--$350$ MeV, while fixing the other parameters by using as input the pion mass and the two-photon decay constant of the neutral pion\footnote{ Particularly, the values $M_\pi=134.9768$ MeV, $\Gamma_{\pi_0\gamma\gamma}=\dfrac{\hbar}{\tau}\dfrac{\Gamma_{\gamma\gamma}}{\Gamma_{tot}}=7.72$ eV, $M_\rho=775.26$ MeV are used \cite{ParticleDataGroup:2022pth}.}.
The inclusion of the vector sector leads to the appearance of an additional four-quark coupling constant $G_2$ which can be fixed in favor of the $\rho$-meson mass, Eq. \ref{VMMassEq}. 
Here one may encounter the absence of confinement. The equation
\begin{align}
k^2+m^2(k^2)=0,
\end{align}	
could be satisfied for some $k^2=-m^2_{\mathrm{pole}}$ and as a result the quark propagator could have singularities. For Gaussian form-factor the type of singularities is poles \cite{Plant:1997jr,Scarpettini:2003fj}. The first pair of them could be real-valued or complex-valued and then there are infinitely many poles in the complex plane. If the pole is real-valued it corresponds to pole mass $m_{\mathrm{pole}}$ and then it is convenient to have $4m^2_{\mathrm{pole}}>M^2_\rho$ in order to define $G_2$ without the generation of an imaginary part. For complex-valued poles the situation is a bit more complicated. Imaginary parts of different poles cancel each other but the real part of the polarization loop has a cusp which should be unphysical. This happens at $\mathrm{M}_{\mathrm{thr}}^2=2\mathrm{Re}(m^2_{\mathrm{pole}}) + 2\sqrt{\mathrm{Re}(m^2_{\mathrm{pole}})^2+\mathrm{Im}(m^2_{\mathrm{pole}})^2}$.
Therefore the region of $m_d$ is restricted to $293-354$ MeV. 

The fitted model parameters are given in the Table \ref{FitModParams}.
In principle $m_c$, $m_d$, $\Lambda$ and $G_1\Lambda^2$ are not independent quantities and are related via the gap equation \eqref{Gap} and four values are given for cross-check.  
In order to prove that the inclusion of vector-sector does not change model parameters drastically, we perform a similar fit without the vector sector and present it in the last three columns in table \ref{FitModParams}. The changes of model parameters after the inclusion of vector--axial-vector sector are only small corrections to model parameters at the 10\% level.

One can see that in the $m_d$ region $293-354$ MeV the variation of $\Lambda$ is around 10 \%, $m_c$ roughly 20 \% and dimensionless ratio $G_2\Lambda^2$ about 30 \%.  
The ratio $G_2/G_1$ is between $-0.08$ and $-0.14$. The calculation with fixed ratios $G_2/G_1=-0.08$, $G_2/G_1=-0.14$ in a wider range of dynamical quark masses is also performed to check the sensitivity of the model to parameter changes.

In the instanton model the effective Lagrangian contains only the scalar--pseudoscalar and the tensor sectors \cite{Shuryak:1983ni} and the relation $|G_2|\ll|G_1|$ is expected.
On the other hand the relation $|G_1|=4|G_2|$ which is valid for large scales can be obtained from one-gluon exchange  \cite{Bijnens:1992uz}. 
In the local NJL model it is found that $|G_2|>|G_1|$ \cite{Volkov:1986zb,Klimt:1989pm} or $|G_1/G_2|$ between $0.78$ and $1.2$ for different fits  \cite{Bijnens:1992uz}. 
It seems that the relation between couplings depends on the model and the scheme used for fixing the model parameters.

The central value is taken for the set of model parameters with $m_d=310$ MeV as it has the maximal $\mathrm{M}_{\mathrm{thr}}$.
For this parameter set one can calculate the properties of the $a_1$ and $f_1$ mesons without any complications and ambiguities due to singularities of the quark propagator. The axial-vector meson mass in the model for this parameter set is found to be $M_{A}=918$ MeV. It is well known that even in the local NJL model the mass of the meson is underestimated \cite{Volkov:1986zb}. 
In the local model, the masses of axial-vector and vector mesons are related by $M_{A}^2=M_\rho^2+6M^2$. 
For the experimental values of meson masses, this relation can be fulfilled only for an unrealistically large mass of the light constituent quark $M\sim400$ MeV.

\section{Two-photon form factor}
\label{FormFactorParams}	

At present, there are only a few experimental data on the form-factor of $1^{++}$ meson decay into two photons.
The L3 Collaboration has studied the reaction
$e^+e^- \to e^+e^-\gamma^\ast\gamma^\ast \to e^+e^-f_1(1285)\to e^+e^- \eta \pi^+\pi^-$
and extracted the form-factor of the $f_1(1285)$ meson transition into pair photons, of which only one is real \cite{Achard:2001uu}. The fitting for the dipole form-factor is
\begin{align}
&\tilde\Gamma_{\gamma\gamma}(A) =  3.5 \pm 0.6 \pm 0.5\,\mathrm{keV},\quad \Lambda_{dip} = 1.04 \pm 0.06 \pm 0.05\,\mathrm{GeV}. \label{ExpL3}
\end{align}

\begin{figure}[t]
\centering
\includegraphics[width=0.49\textwidth]{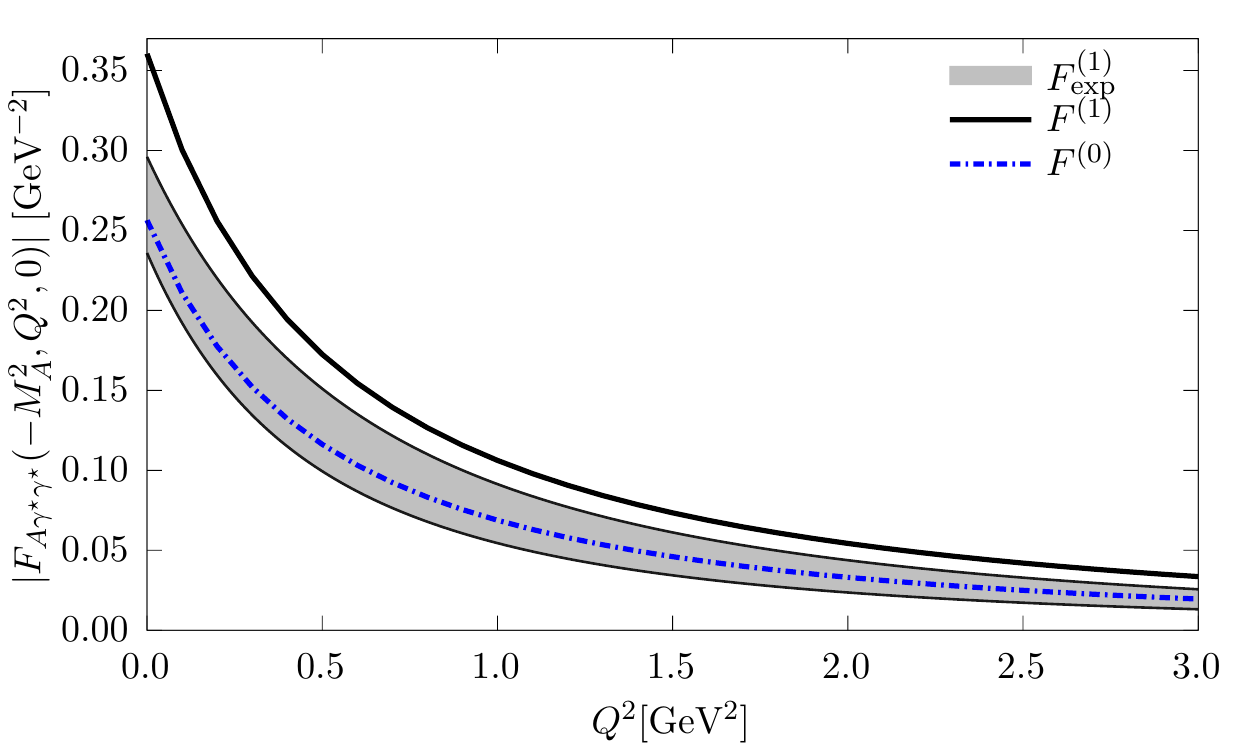}
\includegraphics[width=0.49\textwidth]{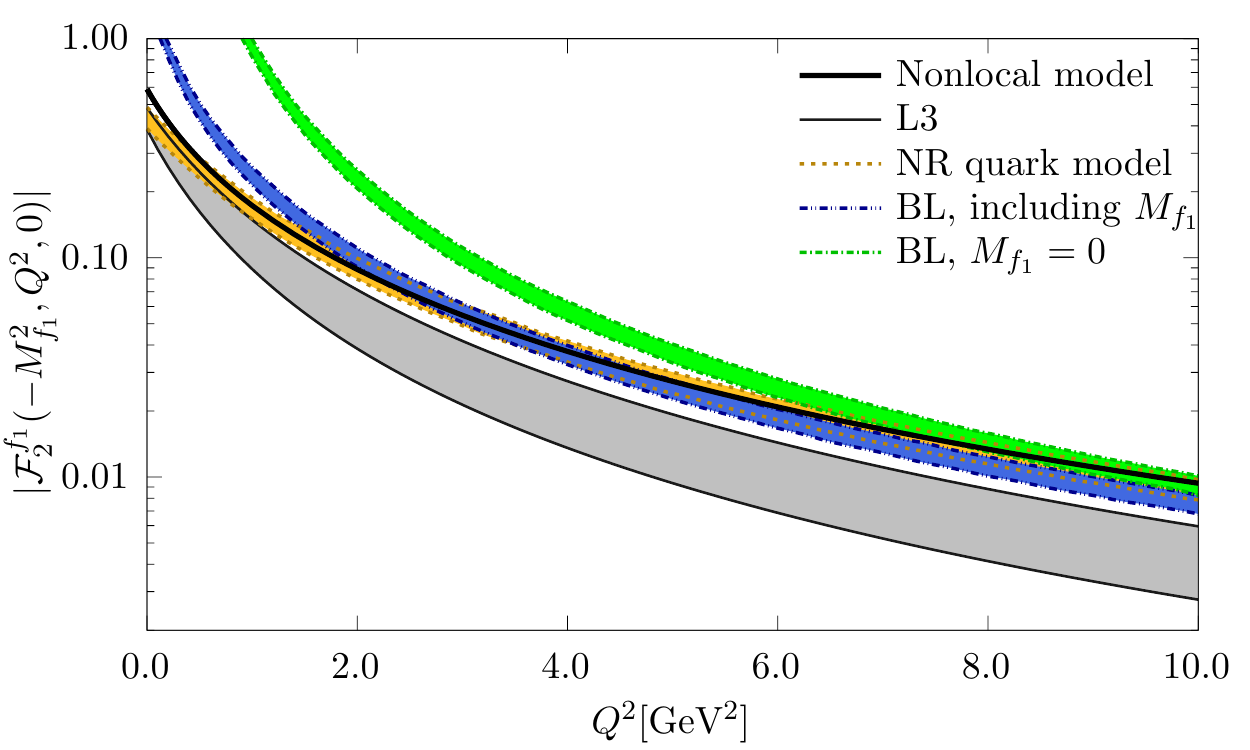}
\caption{
Left: Behaviour of the form-factors \eqref{eq-F0F1} for $f_1(1285)$ on mass-shell. Nonlocal model calculation: black solid line is  $-F^{(1)}_{A\gamma^\ast\gamma^\ast}(M_{A}^2,Q^2,0)$, blue dash-dotted line is $F^{(0)}_{A\gamma^\ast\gamma^\ast}(M_{A}^2,Q^2,0)$. The gray band is the result of the L3 collaboration for $-F^{(1)}_{A\gamma^\ast\gamma^\ast}(M_{A}^2,Q^2,0)$. 
Right: Behaviour of the $\mathcal{F}_2^A$ form-factor \eqref{eqF2} for $f_1(1285)$. The gray band
refers to the L3 result \eqref{ExpL3}, the orange band to the non-relativistic quark model from \cite{Schuler:1997yw}, the green and blue bands to the asymptotic result with or without inclusion of the axial-vector mass \cite{Hoferichter:2020lap}.
}
\label{fig-1}
\end{figure}

We have compared the axial-vector meson form factors\footnote{ 
It is necessary to point out that in Fig. \ref{fig-1} the model value of the axial-vector mass is used for calculations while for the decay width of the particle for phase space the experimental value should be used. 
} ${F}^{(0)}$ and ${F}^{(1)}$  \eqref{eq-F0F1} from above mentioned model calculations to the L3 collaboration result \eqref{ExpL3} in the left part of Fig. \ref{fig-1}. 
The right part of Fig. \ref{fig-1} shows the comparison with the results of the asymptotic transition form-factor $\mathcal{F}_2^A$ from \cite{Hoferichter:2020lap} and the calculation of a non-relativistic quark model \cite{Schuler:1997yw} normalized to L3 data. One can see that when the meson is on the mass shell, the agreement between the model calculation and experimental result is quite reasonable. 

It is also instructive to compare the model with the Brodsky-Lepage scaling result obtained in \cite{Hoferichter:2020lap}. 
The asymptotic constraints for the axial vector transition form-factor were first derived in holographic models in \cite{Leutgeb:2019gbz} and later in the Brodsky-Lepage formalism \cite{Hoferichter:2020lap}.


The results is given in terms of in the average photon virtualities $Q^2$ and the asymmetry parameter $w$ 
\begin{align}
	Q^2=\frac{q_1^2+q_2^2}{2},\qquad w=\frac{q_1^2-q_2^2}{q_1^2+q_2^2}.
\end{align}

At large virtualities the form-factor behaves as
\begin{align}
\mathcal{F}_2^A&=\frac{\mathrm{C}}{Q^4}f_2^A(w),
\end{align}
where $\mathrm{C}$ is sum of different flavour contributions and asymmetry function is
\begin{align}
f^A_2(w)&=\frac{3}{4w^3}\bigg(3-2w+\frac{(3+w)(1-w)}{2w}\log\frac{1-w}{1+w}\bigg). \label{f2aAsymmetry}
\end{align}

\begin{figure}[t]
\centering
\includegraphics[width=0.49\textwidth]{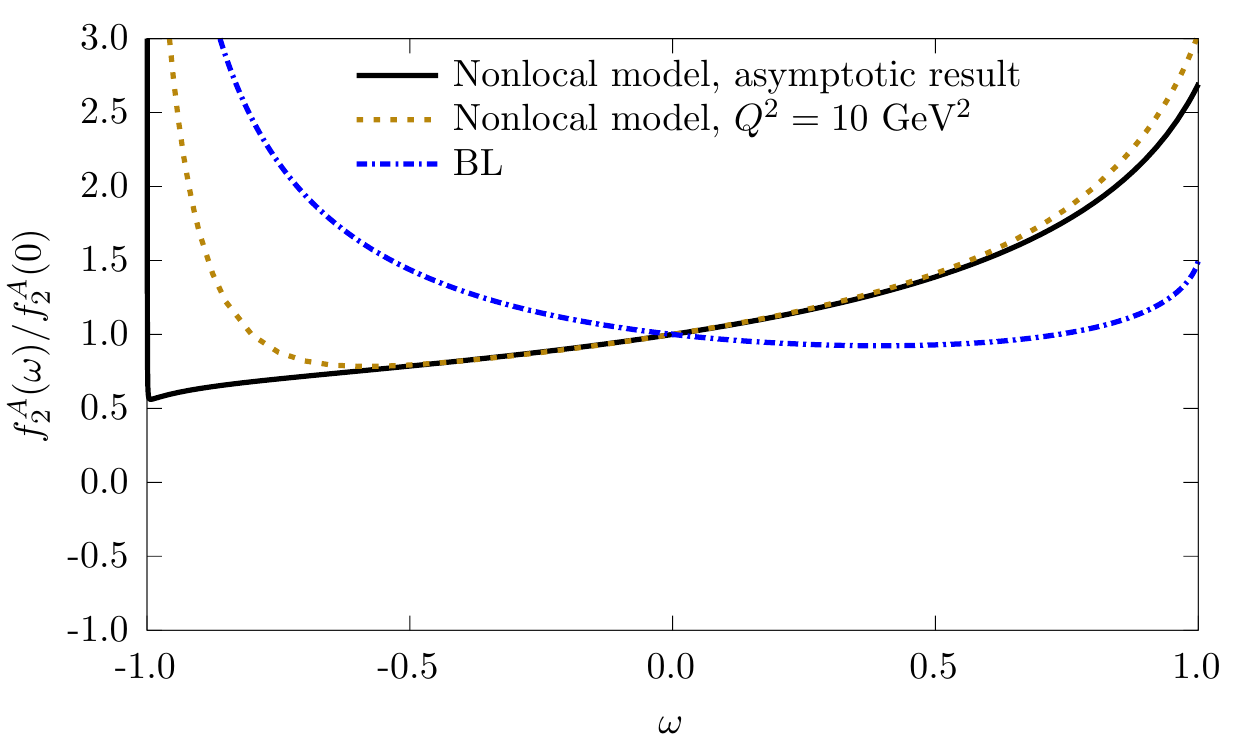}
\caption{
The asymmetry functions $f_2^A(w)/f_2^A(0)$ for the axial-vector meson are shown: the black solid line is the asymptotic result of the nonlocal model calculation, the blue dash-dotted line is the asymptotic result \eqref{f2aAsymmetry} from \cite{Hoferichter:2020lap}, and the yellow dashed line is the generalization of the asymmetry \eqref{f2aAsymmetryGen} for $Q^2=10$ GeV$^2$.
}
\label{fig-Asymmetry}       
\end{figure}

In Fig.\ref{fig-Asymmetry}, the ratio $f_2^A(w)/f_2^A(0)$ calculated from our model is presented. For  $\omega \to -1$, the asymmetry function diverges similarly to \cite{Hoferichter:2020lap}.
It is interesting that this singular behavior for  $\omega \to -1$ is connected with the intermediate $\rho(\omega)-\gamma$ mixing diagrams.
However, in general, the model result does not have a good correspondence to the BL approach. Namely, the behavior near the $-1$ endpoint is not smooth. This could be connected with the properties of the Gaussian form-factor, which leads to too strong suppression. Therefore, it is interesting to consider the generalization of asymmetry functions for finitely large $Q^2$
\begin{align}
\frac{f_2^A(w)}{f_2^A(0)}&= \frac{\mathcal{F}_i^A(-M_A^2,Q^2(1+\omega), Q^2(1-\omega))}{\mathcal{F}_i^A(-M_A^2,Q^2, Q^2)} \label{f2aAsymmetryGen}.
\end{align}
In Fig.\ref{fig-Asymmetry}, the result for $Q^2=10$ GeV$^2$ is presented. One can see that the correspondence with the BL approach becomes slightly better.

\section{Lbl contribution}
\label{lbl}	

\begin{figure}[t]
	\centering
	\begin{center}
		\begin{tabular*}{0.8\columnwidth}{@{}ccccc@{}}
			\raisebox{-0.5\height}{\resizebox{0.25\textwidth}{!}{\includegraphics{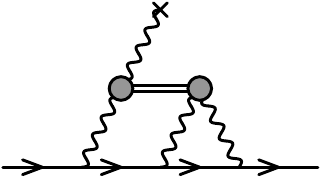}}}&{ } { }&
			\raisebox{-0.5\height}{\resizebox{0.25\textwidth}{!}{\includegraphics{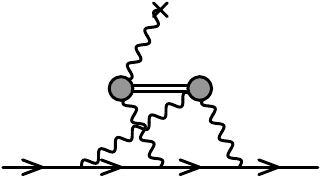}}}&{ } { }&
			\raisebox{-0.5\height}{\resizebox{0.25\textwidth}{!}{\includegraphics{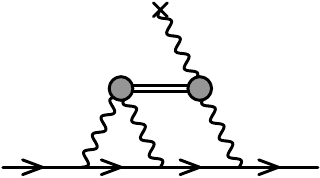}}}\\
			$(a)$&&$(b)$&&$(c)$
		\end{tabular*}
	\end{center}	
	\caption{LbL contribution from intermediate meson exchanges.}
	\label{fig:LbL}
\end{figure}

The LbL contribution to anomalous magnetic moment of the muon is defined by the following projection \cite{Brodsky:1966mv}:
\begin{align}
a_{\mu }^{\mathrm{HLbL}}&=\frac{1}{48m_{\mu }}\mathrm{Tr}\left( (\hat{p}%
+m_{\mu })[\gamma ^{\rho },\gamma ^{\sigma }](\hat{p}+m_{\mu })\mathrm{\Pi }%
_{\rho \sigma }(p,p)\right) , \nonumber \\
& \mathrm{\Pi }_{\rho \sigma }(p^{\prime },p)=-ie^{6}\int \frac{d^{4}q_{1}}{%
	(2\pi )^{4}}\int \frac{d^{4}q_{2}}{(2\pi )^{4}}\frac{1}{%
	q_{1}^{2}q_{2}^{2}(q_{1}+q_{2}-k)^{2}}\times  \label{P4gamProject}\\
& \quad  \times \gamma ^{\mu }\frac{\hat{p}^{\prime }-\hat{q}%
	_{1}+m_{\mu }}{(p^{\prime }-q_{1})^{2}-m_{\mu }^{2}}\gamma ^{\nu }\frac{\hat{%
		p}-\hat{q}_{1}-\hat{q}_{2}+m_{\mu }}{(p-q_{1}-q_{2})^{2}-m_{\mu }^{2}}\gamma
^{\lambda }
\frac{\partial }{\partial k^{\rho }}\mathrm{\Pi }_{\mu \nu
	\lambda \sigma }(q_{1},q_{2},k-q_{1}-q_{2}),  \nonumber
\end{align}%
where $m_{\mu }$ is the muon mass, and
the static limit $k_{\mu }\equiv (p^{\prime }-p)_{\mu }\rightarrow 0$ is implied. The four-rank polarization tensor $\mathrm{\Pi }_{\mu \nu\lambda \sigma }$ is saturated by resonances, as shown in Fig. \ref{fig:LbL}. 

For pseudoscalar mesons the polarization tensor reads
\begin{align}
\mathrm{\Pi}^{\mu\nu\lambda\rho}(q_1,q_2,q_3)
&=i 
\Delta^{\mu\nu}_{P}(q_1+q_2,q_1,q_2) 
\frac{1}{(q_1+q_2)^2-M_P^2}
\Delta^{\lambda\rho}_{P}(q_1+q_2,q_3,q_4)
+\notag\\
&+i
\Delta^{\mu\rho}_{P}(q_2+q_3,q_1,q_4)
\frac{1}{(q_2+q_3)^2-M_P^2}
\Delta^{\nu\lambda }_{P}(q_2+q_3,q_2,q_3        )
+\notag\\
&+i
\Delta^{\mu\lambda}_{P}(q_1+q_3,q_1,q_3        )
\frac{1}{(q_1+q_3)^2-M_P^2}
\Delta^{\nu\rho    }_{P}(q_1+q_3,q_2,q_4).\label{PiLblExpr}
\end{align}

The expression for other mesons can be rewritten in the same way with the corresponding changes of form-factors and propagators. For example, the first line in \eqref{PiLblExpr} for the axial-vector meson is:
\begin{align}
i\Delta_{A,\alpha}^{\mu\nu}(q_1+q_2,q_1,q_2) 
\bigg(\mathrm{D}^{\mathrm{T};R}_M((q_1+q_2)^2)\mathrm{P}^{T;\alpha\beta}_{q_1+q_2}+\mathrm{D}^{\mathrm{L};R}_M((q_1+q_2)^2)\mathrm{P}^{L;\alpha\beta}_{q_1+q_2}\bigg)
	\Delta_{A,\beta}^{\lambda\rho}(q_1+q_2,q_3,q_4).
\end{align}  

Then, by averaging over the direction of the muon momentum, the result for $a_{\mu }^{\mathrm{HLbL}}$ becomes a three-dimensional integral with the radial integration variables $Q_{1},Q_{2}$ and the angular variable \cite{Dorokhov:2012qa,Jegerlehner:2009ry,Jegerlehner:2017gek,Colangelo:2015ama,Colangelo:2017fiz}.
Since the additional momentum in the meson-two-photon vertex is the meson one which is already present in meson propagator the averaging procedure is not changed. 

After integrating over the angular variable, the LbL contribution can be written in the form \cite{Dorokhov:2015psa}
\begin{equation}
a_{\mu}^{\mathrm{LbL}}=\int\limits_{0}^{\infty}dQ_{1}\int\limits_{0}^{\infty
}dQ_{2}\,\,\rho(Q_{1},Q_{2}), \label{aLbL4}%
\end{equation}
where $\rho(Q_{1},Q_{2})$ is the density function for the  contribution to $g-2$. 
We cross-checked that our program reproduces the expression for the projectors proposed in \cite{Jegerlehner:2009ry} as well as the numerical results for the pion pole \cite{Knecht:2001qf}.

\begin{figure}[t]
\centering
\includegraphics[width=0.49\textwidth]{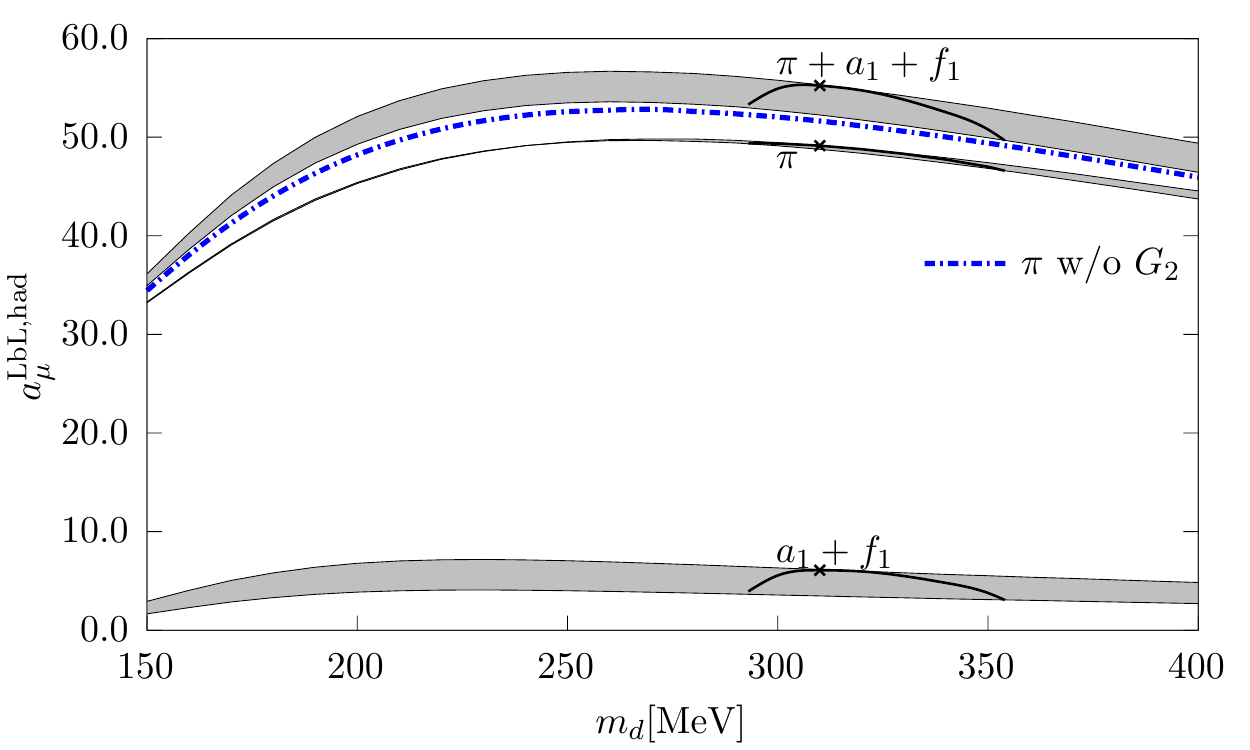}
\includegraphics[width=0.49\textwidth]{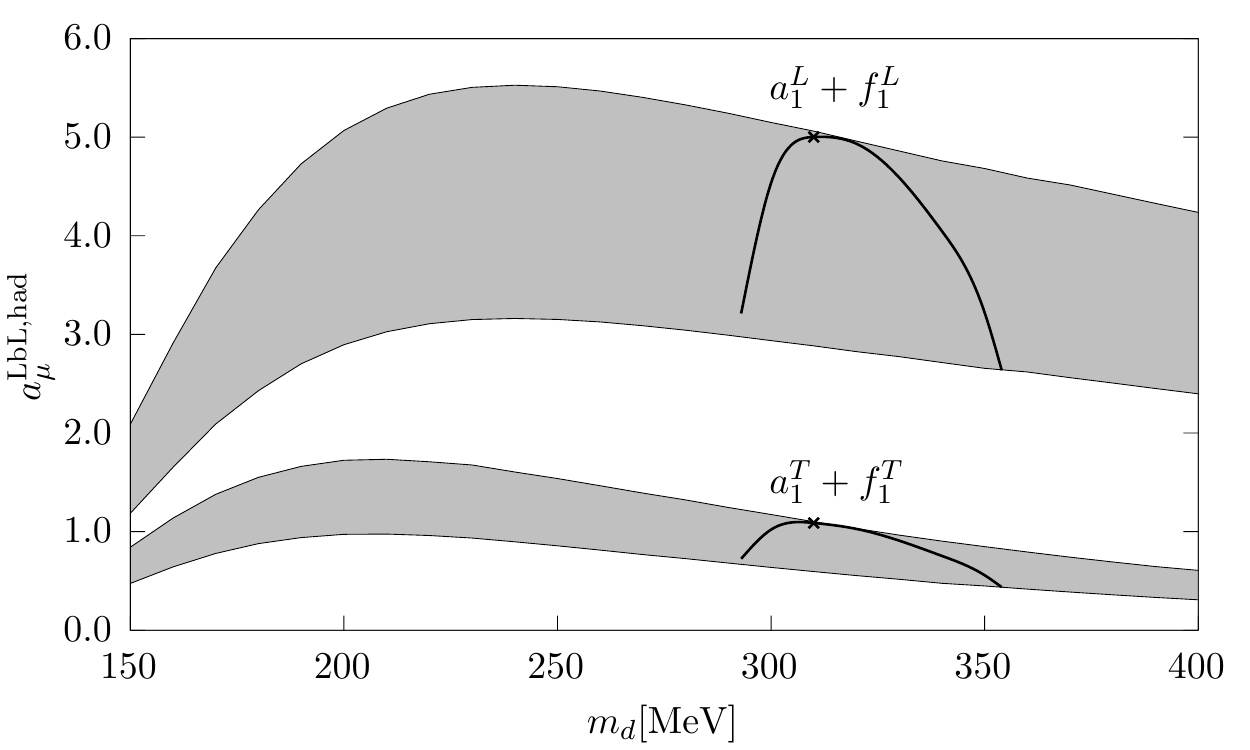}
\caption{ (in units $10^{-11}$)
Left: The LbL contribution to the muon AMM from the $\pi$, $a_1+f_1$ and $\pi+a_1+f_1$ exchanges for different parameterizations as a function of the dynamical quark mass. The shaded area corresponds to the region between fixing $G_1/G_2=-0.08$ and $G_1/G_2=-0.14$. The black solid lines correspond to the parameterization where $G_2$ is fixed in order to reproduce the physical value of $\rho$-meson mass. The cross corresponds to the result with $m_d=310$ MeV. The blue dashed-dotted line is the pion contribution in the model without the vector sector \cite{Dorokhov:2011zf}. Right: The separate contribution of the longitudinal and transverse components of $a_1+f_1$ mesons for different parameterizations as a function of the dynamical quark mass. }
	\label{fig-AllLbl}       
\end{figure}	
\section{Result}
\label{Result}	

The LbL contribution of $\pi$ with the $\pi-a_1$ mixed component and the $a_1+f_1$ mesons is shown in the left part of Fig. \ref{fig-AllLbl}. Recall that ideal mixing is implemented for $a_1$ and $f_1$ mesons. The partial contributions of the transverse and longitudinal components of the axial meson are shown in the right part of Fig. \ref{fig-AllLbl}. Surprisingly, most of the contribution comes from the longitudinal part.

The analysis of the axial meson contribution is complicated by the mixing between $a_1$ meson and pion, which leads to a change in the pion properties. Therefore, it is necessary to refit the model parameters before performing the comparison.
However, after refitting the model parameters, the pion LbL contribution is also changed.
Hence, it seems that it is more meaningful to subtract the result of the $\pi$ contribution in the model without-vector-sector from the full $\pi+a_1+f_1$ contribution in the model with-vector-sector, for a given $m_d$.
In this way, the (additive) axial contribution can be extracted (from the full $\pi+a_1+f_1$ contribution).    

It is helpful to consider a set of model parameters with $m_d=310$ MeV, since this set is used as a central point for estimations. It can be shown that pion LbL contribution in the model without the vector sector is  $51.6$, $\pi+a_1+f_1$ LbL contribution  in the model with the vector sector is $55.2$, so the $a_1+f_1$ contribution is $3.6$. 


The error band is estimated as the difference between the highest and lowest values of the contribution, i.e. for $G_2/G_1=-0.14$ and $G_2/G_1=-0.08$, respectively.


Therefore, the estimated $a_1+f_1$ contribution is $3.6 \pm 1.8$.
For rough estimations of the strange quark contributions, the values of the current and dynamical mass of the strange quark can be taken from \cite{Dorokhov:2015psa} (using interpolation for values between $m_{d}=305$ and $320$ MeV) : $m_{c,s}=239.9$ MeV, $m_{d,s}=418.2$ MeV. $G_2$ and $\Lambda$ are taken from the set with $m_d=310$ MeV. As a result, in ideal mixing, the  $f'_1$ contribution is found to be small (from T and L modes):
\begin{align}
0.004+0.039=0.041 \label{fprimecont}
\end{align}
and therefore does not change the final answer for $a_1+f_1$ contribution.

\section{Discussion}
\label{Discussion}	

\begin{table}
\centering
\caption{Contribution to $a_\mu$ from axial-vector meson exchanges (in $10^{-11}$). The $a_1$(1260), $f_1$(1285) and $f_1$(1420) mesons are denoted as $a_1,f_1,f_1'$. When possible, the partial contributions of the mesons are given in the note field in brackets. T and L denote transverse and longitudinal modes, respectively. }
\label{tab-1}       
\begin{tabular}{|c|c|p{0.6\textwidth}|}   \hline
Ref. & Contribution & Note  \\
\hline
\cite{Bijnens:1995cc,Bijnens:1995xf} & $2.5 \pm 1$ & ENJL model    \\
\cite{Hayakawa:1996ki,Hayakawa:1997rq} & $1.74\pm0.35$  & ENJL model  \\
\cite{Melnikov:2003xd}& $22\pm 5$  & Resonance+OPE, partial contributions $a_1,f_1,f_1'$: $\{5.7 + 15.6 +0.8\}$  \\
\cite{Pauk:2014rta} & $6.4 \pm 2$ & Resonance, dipole FF from L3, partial contributions $f_1,f_1'$: $\{5 + 1.4\}$   \\
\cite{Jegerlehner:2017gek} & $7.55 \pm 2.71$ &Resonance+OPE, partial contributions $a_1,f_1,f_1'$: $\{1.89 + 5.19 + 0.47\}$   \\
\cite{Cappiello:2019hwh}& ${28}$ & AdS/CFT, tower of resonances in channels $a_1,f_1,f_1'$, partial T and L modes: $\{4+4+6\}$ and $\{4+4+6\}$    \\
\cite{Leutgeb:2019gbz} 	& ${ 22\pm5}$ & AdS/CFT, tower of resonances in channels $a_1,f_1,f_1'$,  partial T and L modes: $\{9\}$ and  $\{13\}$ \\
\cite{Roig:2019reh} 	& $ 0.8^{+3.5}_{-0.1}$ & Resonance chiral theory, $a_1,f_1,f_1'$\\
\cite{Leutgeb:2022lqw} 	& ${27.8}$& AdS/CFT with the U(1)$_A$ anomaly, v1(OPE) fit, partial contributions $a_1,f_1,f_1'$: $\{7.8 + 5.71 + 14.3\}$ \\
& ${25}$& 
v1(F$_\rho$-fit), partial contributions $a_1,f_1,f_1'$: $\{7.1 + 4.34 + 13.6\}$\\
\hline
\cite{Prades:2009tw}& $15\pm 10$& ``Glasgow consensus''   \\
\cite{Aoyama:2020ynm}& $6\pm6$& ``White paper'' 2020 \\
	\hline
	This work & $3.6 \pm 1.8$ &Ratio of T and L modes contribution: T/L$=\{ 0.22-0.43\}$. Estimation for $f'_1$ is $0.041$\\
	\hline
\end{tabular}
\end{table}

In the nonlocal chiral quark model, the contribution of $a_1(1260)$ and $f_1(1285)$  in the case of ideal mixing is estimated to be $(3.6 \pm 1.8) \cdot 10^{-11}$, which includes the full kinematic dependence of the vertices, i.e., the off-shell effects for the mesonic bound state, the reevaluation of the pion contribution resulting from $\rho-\gamma$ and $\pi-a_1$  mixing, and the refitting of the model parameters. 
Moreover, it is normalized to the model without-vector-sector, so that the contributions from different parts could be added, thus making it possible for further comparison with the other models.

The axial-vector contribution is presented in Table \ref{tab-1} together with other estimations.

At first glance, our result is comparable in size to that of
\begin{itemize}
\item	
early estimates in the NJL-like models \cite{Bijnens:1995cc,Bijnens:1995xf,Hayakawa:1996ki,Hayakawa:1997rq}, 
\item
using form-factors from L3 collaboration data \cite{Pauk:2014rta},	
\item
resonances + OPE restriction \cite{Jegerlehner:2017gek},
\item
resonances chiral theory  \cite{Roig:2019reh}.
\end{itemize}

Our result is lower than 
\begin{itemize}
	\item	
Melnikov-Vainshtein estimates \cite{Melnikov:2003xd},
	\item	
AdS/CFT calculations including the tower of resonances \cite{Cappiello:2019hwh} and \cite{Leutgeb:2019gbz}, 
	\item	
AdS/CFT calculations with the U(1)$_A$ anomaly \cite{Leutgeb:2022lqw}. 
\end{itemize}

Our result is compatible with the ``White paper'' 2020  estimate \cite{Aoyama:2020ynm}. 

However, it should be pointed out that a detailed understanding of the axial-vector contribution a quite complicated task. The following points which are should be discussed:
\begin{itemize}
\item dependence of the transition form-factor on meson virtuality,
\item contribution of the longitudinal mode,
\item the short-distance constraint for LbL amplitude,
\item correspondence with the whole result.
\end{itemize}

\begin{figure}[t]
\centering
\includegraphics[width=0.49\textwidth]{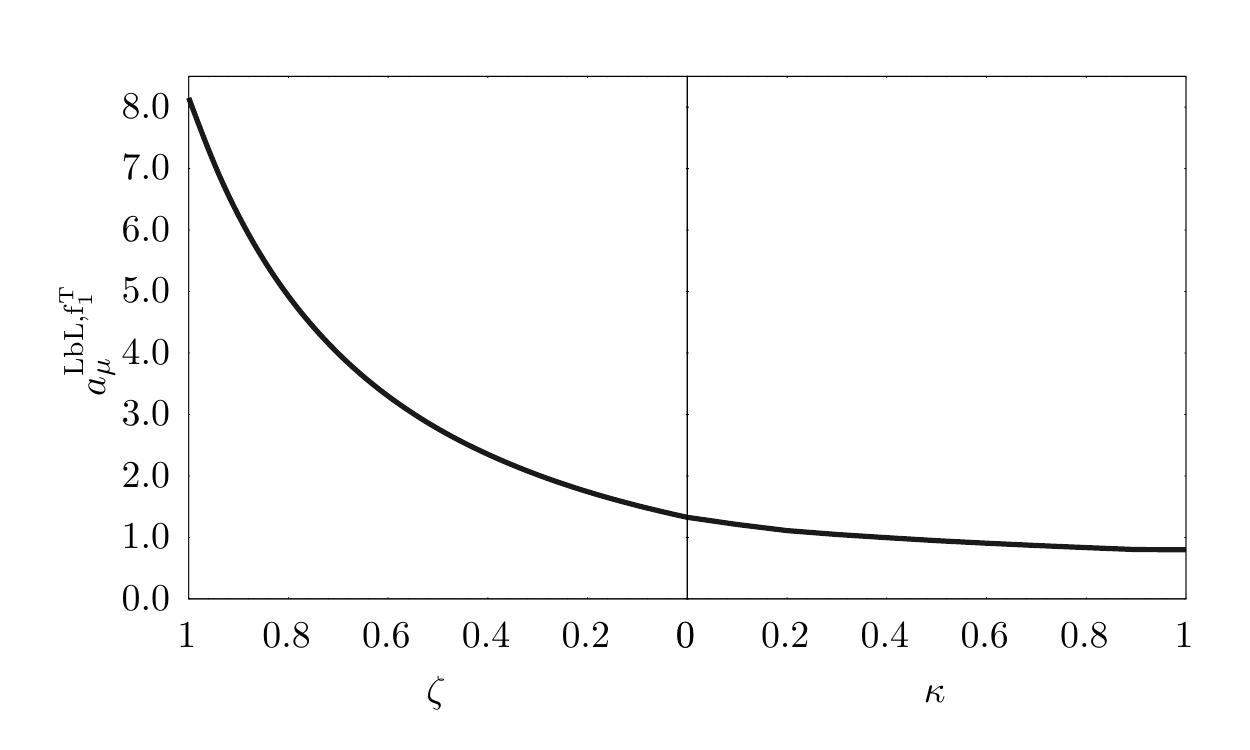}
\caption{
The contribution of the transverse component of $f_1$ meson to the anomalous magnetic moment at different values of the artificial momentum $P^2$ where $\zeta=- M_A^2/P^2$ and $\kappa= Q_1^2/P^2$. 
}
\label{fig-MultiV2}       
\end{figure}

In our model, the transition form-factor depends not only on the virtuality of photons but also on that of meson. This dependence results from the fact that mesons are quark-antiquark bound states. It leads to suppression of the LbL contribution due to the virtuality of the meson and in \cite{Dorokhov:2011zf} it is  discussed for $\pi$, $\eta$, $\eta'$ mesons. Similar situation should occur in the local NJL model \cite{Bijnens:1995xf} or the DSE approach \cite{Goecke:2010if}. There are approaches based on phenomenological information from experiments, i.e. making use of the form-factor when the meson is on the mass shell \cite{Pauk:2014rta,Jegerlehner:2017gek}.
In the model, one can investigate the on-mass-shell case in the following way. Firstly, the part of a diagram that contains the form-factors in LbL can be rewritten in the following form in Euclidean space:
\begin{align}
	A_i(P^2,(Q_1+Q_2)^2,Q_2^2) 
	\frac{1}{Q_1^2+M_A^2}
	A_j(P^2,Q_1^2,0),
\end{align}
where $A_i$ and $A_j$ are the form-factor functions, the additional 
artificial momentum $P^2$ is introduced in order to investigate the dependence of the transition form-factors on meson virtuality. The two limiting cases are:
\begin{itemize}
\item
$P^2=-M_A^2$ corresponds to the case where the form-factors are taken when the meson is on-shell;
\item
$P^2=Q_1^2$ corresponds to the case where the dependence of form-factors on meson virtuality is taken into account.
\end{itemize}
In order to connect these limits smoothly, we consider the following two-step transition. First, fix the artificial momentum at $P^2=-\zeta M_A^2$ and $\zeta$ is decreased from $1$ to $0$. Then the artificial momentum is taken to be $P^2=\kappa Q_1^2$ where $\kappa$ is increased from 0 to 1. The same procedure is performed for three possible diagrams, with appropriate changes of $P^2$. In Fig. \ref{fig-MultiV2}, the LbL contribution for the $f_1$ meson is presented for this two-step transition.  

In the limit $P^2=-M_A^2$ one can estimate the axial-vector contributions of the transverse modes for $a_1$, $f_1$ and $f_1'$ mesons with approximation $g_{a_1}=g_{f_1}=g_{f_1'}$ and PDG values of meson masses for two scenarios: ideal mixing $\theta=35.3^\circ$ and $\theta=62^\circ$ from L3 experiment\cite{Achard:2001uu}. The results for the partial contributions from $a_1$, $f_1$, $f_1'$ are 
\begin{itemize}
	\item ideal mixing: $1.95+5.08+0.004=7.03$,
	\item L3, $\theta=62^\circ$: $1.95+3.63+1.24=6.82$.
\end{itemize}
As a result, one can see that for the on-mass-shell form-factors, the result obtained in the paper is not far from \cite{Pauk:2014rta,Jegerlehner:2017gek}.
The strange quark contribution is also suppressed as in \eqref{fprimecont}. The source of this suppression could be understand using the local NJL model: the $A_i$ functions at zero virtuality are proportional to $1/M^2$, where $M$ is the constituent mass and the ratio of strange to nonstrange masses is $r_{s/l}=M_s/M_u\sim1.6$ \cite{Volkov:1986zb,Bijnens:1995xf}. As a result even if we suppose that most of the changes in $A_i$ for nonstrange and strange quarks are from the constituent quark mass values, 
the $f_1'$ contribution will be suppressed by order $\frac{2}{9}r_{s/l}^{-4}\sim 30$ in comparison with $a_1$. Then,
one can naively invert the ratio of $a_1$ and $f'$ contributions in the ideal mixing case in order to find the effective ratio of masses $r_{s/l}^{\mathrm{eff}}$
\begin{align}
	r_{s/l}^{\mathrm{eff}}=
	\left(\frac{1.95}{0.004}\cdot\frac{2}{9}\cdot\left(\frac{1230}{1420}\right)^2\right)^{1/4}\sim 3.
\end{align}
In the nonlocal model, the ratio of $a_1$ and $f'$ contributions is suppressed even stronger due to the momentum dependence since at the  zero virtuality, the ratio of masses is somewhere between $m_{s}(0)/m_{u}(0)\approx 2$ and $m_{s}(\infty)/m_{u}(\infty)=m_{c,s}/m_{c,u}\approx 31$. From the equation $r_{s/l}^{\mathrm{eff}}= m_{s}(\mu^2_{\mathrm {eff}})/m_{u}(\mu^2_{\mathrm {eff}})\sim3$, one can estimate that such ratio could be reached when $\mu_{\mathrm {eff}}\sim 0.7$ GeV. 

Another important contribution has to do with the longitudinal modes. By construction, the spin-1 field contains part that corresponds to spin 0. As a result, this part of the field mixes with the pseudoscalar one.
In the case of $a_1$ meson, the corresponding particle is pion and therefore the interplay of the longitudinal spin-0 component becomes important.
In the nonlocal model, most of the contribution comes from the L mode: the $T/L$ ratio is $0.22$ for uncorrected data and $0.43$ after normalization of the pion contribution to the model without vector sector. 
In \cite{Cappiello:2019hwh}, using AdS/CFT correspondence to hadronic physics, the T and L contributions of the axial-vector towers of resonances are estimated to be of the same size.
In \cite{Leutgeb:2019gbz}, it is discussed that results of \cite{Cappiello:2019hwh} 
roughly correspond to the HW2 holographic model results of \cite{Leutgeb:2019gbz} with different sets of parameters. In \cite{Leutgeb:2019gbz}, the $T/L$ ratios quoted for the HW1, HW2 models and an extrapolation to match the L3 data for $f_1$ and $f_1'$
are $17.4/23.2\approx0.75$, $12.0/16.6\approx0.72$ and $9/13\approx0.69$, correspondingly.    
It is interesting that the only T-mode contribution in the nonlocal model is similar to that obtained in the resonance chiral theory \cite{Roig:2019reh}.

The short distance constraint for the LbL amplitude in the nonlocal model is connected with the quark loop contribution, which is discussed in more details  \cite{Dorokhov:2011zf} (see also \cite{Dorokhov:2005hw,Dorokhov:2005pg}). In the model, the quark-loop contribution provides the correct asymptotic, while the meson contributions do not violate the OPE results due to the dependence of form-factors on meson virtuality. On the other hand, for the resonances, the OPE restrictions can be achieved by an infinite tower of resonances \cite{Cappiello:2019hwh,Leutgeb:2019gbz,Masjuan:2020jsf}.
For example, in the five-dimensional model where the fifth dimension is later integrated out and which mimics QCD in the large-$N_c$ limit \cite{Cappiello:2019hwh} the short distance constraint is achieved by the infinite tower of the axial-vector contribution of $a_1(1260)$, $f_1(1285)$ and $f_1^*(1420)$. 
Similarly, in \cite{Leutgeb:2019gbz}, the axial-vector meson contributions arising from the Chern-Simons action in the holographic QCD models are estimated and it is found that the infinite tower of axial-vector mesons leads to correct behavior at short distances. 
In this sense, from quark-hadron duality arguments, the quark loop represents the contribution of an infinite number of exited states since only the ground states are explicitly included in our model. 

The contribution of $u$, $d$, $s$ quark loop is estimated in \cite{Dorokhov:2015psa} in the nonlocal model without spin-1 fields. The total result
\begin{equation}
	a_{\mu}^{\mathrm{HLbL, N\chi QM}} =168\pm 12.5, \label{aNchiQM}%
\end{equation}
and the partial contributions from pseudoscalar($\pi^{0}$, $\eta$, $\eta^{\prime}$) mesons $58.5\pm8.7$, the scalar ($\sigma$, $a_{0}(980)$, $f_{0}(980)$) mesons
$3.4\pm4.8$ and the quark loop $110\pm9$. 
With the addition of the axial-vector mesons, the central value becomes $171.6$.
The number quoted for the total LbL contribution in ``White paper'' 2020 \cite{Aoyama:2020ynm} and ``Snowmass 2021''  \cite{Colangelo:2022jxc} from phenomenology is $92\pm19$ while recent lattice results are $109.6\pm15.9$ \cite{Chao:2022xzg} and $124.7\pm14.9$ \cite{Blum:2023vlm}.
Therefore, the question of double-counting in the model arises, since both contribution from mesons and quark alone is not far from the ``White paper'' or lattice values. However, our calculations with the inclusion of the vector–axial-vector sector are not yet complete: while the model parameters refitted with the inclusion of the vector sector are only mildly changed, the effect of $\rho-\omega$ mixing influence on the quark loop contribution is not yet properly accounted in quark loop contribution.

The diagrams of the type shown in Fig.\ref{fig:GGaqq}(d), Fig.\ref{fig:GGaGaqq}(c,d) and similarly with three or four photons should be taken into account. The very rough estimation in the local NJL model of this effect can be done based on the value quoted in \cite{Bijnens:1995xf} with the expression for the LbL contribution for the constituent quark loop \cite{Pivovarov:2001mw,Greynat:2012ww}. This analytical expression is the contribution
of a fermion to the order $(m_\mu/M)^4$ \cite{Laporta:1992pa} with appropriate factors. 
The values of constituent quark mass from  \cite{Bijnens:1995xf} ($M_u=275$ MeV, $M_s=427$ MeV), result for the quark loop contribution using the expression is $62$, while the value quoted in \cite{Bijnens:1995xf} is $21(3)$, signals that there is strong suppression due to $\rho-\omega$ mixing.
On the other hand the effects of dressing is estimated to be 5\% in DSE approach \cite{Goecke:2012qm}. 

Furthermore, one can estimate what the constituent masses should be to give the nonstrange and strange contribution, of the size of the nonlocal model one.
By using the values $u+d$ quark contribution of $99.8$ for $m_{d,u}=310$ MeV, and strange contribution of $1.89$\footnote{Using interpolation for values between $m_{d,u}=305$ and $320$ MeV.} in the nonlocal model the constituent quark masses $M_u=209$ MeV and $M_s=383$ MeV  can be obtained. One can use these values to check reasonability of scale in momentum dependent mass as  $m_u(\mu_{\mathrm {eff}}^2)=M_u$ and  $m_s(\mu_{\mathrm {eff}}^2)=M_s$. The corresponding results for the momentum scales from these equations are more or less reasonable  $\mu_{\mathrm {eff}}\approx 0.47$ GeV and $\mu_{\mathrm {eff}}\approx 0.73$ GeV.

Apart from the suppresion due to $\rho-\omega$ mixing, another source for the decreasing of the phenomenological value are the terms formally suppressed by $1/N_c$, i.e., the  contribution from pion and kaon loops. One can find the estimation of these contributions in ``White paper''\footnote{The recent DSE approach calculations gave somewhat similar value $-16.1\pm0.2$ \cite{Miramontes:2021exi}.} $-16.4\pm0.2$. In the nonlocal model estimations sub-leading corrections can be done with help of $1/N_c$ expansion \cite{Plant:2000ty,Radzhabov:2010dd}.

\section{Conclusions}
\label{Conclusions}	

The LbL contribution of axial-vector mesons to the anomalous magnetic moment of the muon is estimated in the framework of the nonlocal quark model, which contains only ground state mesons.
The inclusion of axial-vector particles leads to mixing with pseudoscalar states, the vertices of interaction with photons are dressed by vector particles. 
The model parameters are refitted.
In order to obtain the final result, the contribution from $\pi+a_1+f_1$ in the model with the vector sector is normalized to the pion result in the model without the vector sector. 

In the nonlocal quark model, the meson transition form factors depend not only on the photon virtuality, but also on the meson virtuality. This leads to a strong suppression, and the resulting value for the $a_1+f_1$ LbL contribution is $(3.6 \pm 1.8) \cdot 10^{-11}$  (estimation for the $f'_1$ is $0.041$). Most of the contribution comes from the longitudinal mode.
In the future, we plan to re-estimate the influence of the vector–axial-vector sector on the quark loop contribution and to extend our calculations to include sub-leading $1/N_c$ corrections. In fact, the presence of the quark loop is the main difference between models with quark degrees of freedom and that with only mesonic degrees of freedom or the dispersive approach. Up to now, it is not clear how to relate these calculations since the mesonic contributions are present in both approaches, while the quark loop is attributed only to the quark models.

A clear understanding of the relation between the quark model calculations and the dispersive approach needs an extension of the nonlocal model beyond leading $1/N_c$ order, at which the mesonic bound states appear in the loops, therefore one can directly compare them with the dispersive approach.
The first step in this direction is made in \cite{Radzhabov:2010dd} where the nonlocal chiral quark model is extended beyond mean field.

The authors thank the \boxed{\rm{A.E.~Dorokhov}}, our friend and scientist, with whom we started this work. The authors are grateful to Baiyang Zhang for careful reading of our manuscript and useful remarks. 

Diagrams are drawn with the help of the \verb|feyn.gle| package \cite{Grozin:2022fde}.
This work is supported by the Russian Science Foundation (Grant No. RSF 23-22-00041).

\bibliographystyle{apsrev4-2}

\end{document}